\documentclass[11pt,a4paper]{article}%
\usepackage{amsfonts}
\usepackage{amsmath}
\usepackage{amssymb}
\usepackage{graphicx}
\usepackage{geometry}%
\providecommand{\U}[1]{\protect\rule{.1in}{.1in}}
\newtheorem{theorem}{Theorem}

\newtheorem{corollary}{Corollary}

\newtheorem{notation}{Notation}

\newtheorem{proposition}{Proposition}
\newtheorem{remark}{Remark}

\makeatletter
\def \@removefromreset#1#2{\let \@tempb \@elt
\def \@tempa#1{@&#1}\expandafter \let \csname @*#1*\endcsname \@tempa
\def \@elt##1{\expandafter \ifx \csname @*##1*\endcsname \@tempa \else
\noexpand \@elt{##1}\fi}     \expandafter \edef \csname cl@#2\endcsname{\csname cl@#2\endcsname}     \let \@elt \@tempb
\expandafter \let \csname @*#1*\endcsname \@undefined}

\@removefromreset{equation}{section}

\@removefromreset{theorem}{section}
\makeatother
\begin{document}

\title{The Bloch vectors formalism for a finite-dimensional quantum system }
\author{Elena R. Loubenets$^{1,2}$ and Maxim S. Kulakov$^{1}$\\$^{1}$National Research University Higher School of Economics, \\Moscow 101000, Russia\\$^{2}$Steklov Mathematical Institute of Russian Academy of Sciences, \\Moscow 119991, Russia}
\maketitle

\begin{abstract}
In the present article, we consistently develop the main issues of the Bloch
vectors formalism for an arbitrary finite-dimensional quantum system. In the
frame of this formalism, qudit states and their evolution in time, qudit
observables and their expectations, entanglement and nonlocality, etc. are
expressed in terms of the Bloch vectors -- the vectors in the Euclidean space
$\mathbb{R}^{d^{2}-1}$, arising under decompositions of observables and states
in different operator bases. Within this formalism, we specify for all
$d\geq2$ the set of Bloch vectors of traceless qudit observables and describe
its properties; also, find for the sets of the Bloch vectors of qudit states,
pure and mixed, the new compact expressions in terms of the operator norms
that explicitly reveal the general properties of these sets and have the
unified form for all $d\geq2$. For the sets of the Bloch vectors of qudit
states under the generalized Gell-Mann representation, these general
properties cannot be analytically extracted from the known equivalent
specifications of these sets via the system of algebraic equations. We derive
the general equations describing the time evolution of the Bloch vector of a
qudit state if a qudit system is isolated and if it is open and find for both
cases the main properties of the Bloch vector evolution in time. For a pure
bipartite state of a dimension $d_{1}\times d_{2}$, we quantify its
entanglement via the characteristics of the Bloch vectors for its reduced
states. The introduced general formalism is important both for the theoretical
analysis of quantum system properties and for quantum applications, in
particular, for optimal quantum control, since, for systems where states are
described by vectors in the Euclidean space, the methods of optimal control,
analytical and numerical, are well developed.

\end{abstract}

\section{Introduction}

For qubit states and qubit observables, the formalism of Bloch vectors
(coherence vectors) is well developed \cite{1,2,3,4} and is widely used in
many quantum information fields, for example, in quantum computation \cite{3}.
This is not, however, the case for a qudit system of an arbitrary dimension
$d\geq2.$

For $d\geq3,$ in the literature, mostly the properties of the Bloch vectors of
qudit states under the generalized Gell-Mann representation
\cite{5,6,7,8,9,10,14} and the problems of their visualization
\cite{11,12,13,15} have been analyzed.

It is, however, important to develop a general formalism, where, for a
finite-dimensional quantum system of an arbitrary dimension $d\geq2$, not only
its states but also its observables, its evolution in time, the entanglement
and nonlocality, etc. would be described in terms of the Bloch vectors -- the
vectors in the Euclidean space $\mathbb{R}^{d^{2}-1}$, arising under
decompositions of observables and states in different operator bases. The new
results in this direction have been recently presented in \cite{16,17,18}.

The development of this general formalism is important both for the
theoretical analysis of quantum system properties and for quantum
applications, in particular, for optimal quantum control, since, for systems,
which states are described by vectors in the Euclidean space, the methods of
optimal control, analytical and numerical, are well known.

In the present article, we consistently formalize (Sections 2, 3) and classify
the main properties of Bloch-like representations for linear operators on a
finite-dimensional complex Hilbert space. This allows us: (a) to specify
(Section 4) the geometry properties of the set of Bloch vectors for all
traceless qudit observables; (b) to find (Section 5) for the sets of the Bloch
vectors of all qudit states, pure and mixed, the new compact expressions in
terms of the operator norms, which explicitly reveal the general geometry
properties of these sets and have the unified form for all $d\geq2;$ (c) to
derive (Sections 6) the images in the Euclidean space of the Liouville--von
Neumann equation and the Lindblad master equation and to find for the Bloch
vector of a qudit state the main properties of its time evolution if a qudit
system is isolated and if it is open; (d) to quantify (Section 7) the
entanglement of a pure bipartite state in terms of the Bloch vectors for its
reduced states.

\section{Operator bases}

Let $\mathcal{H}_{d}$ be a complex Hilbert space of a finite dimension
$d\geq2$ and $\mathcal{L}_{d}$ denote the vector space of all linear operators
$X$ on $\mathcal{H}_{d}$ equipped with the scalar product
\begin{equation}
\langle X_{i},X_{j}\rangle_{\mathcal{L}_{d}}:=\mathrm{tr}\left(  X_{i}^{\dag
}X_{j}\right)  . \label{1}%
\end{equation}
Denote by%
\begin{align}
\mathfrak{B}_{\Upsilon_{d}}  &  :=\left\{  \mathbb{I}_{d},\text{ }\Upsilon
_{d}^{(k)}\in\mathcal{L}_{d},\text{ \ }k=1,...,(d^{2}-1)\right\}  ,\label{2}\\
\Upsilon_{d}^{(k)}  &  =\left(  \Upsilon_{d}^{(k)}\right)  ^{\dagger}%
\neq0,\ \ \ \mathrm{tr}\left(  \Upsilon_{d}^{(k)}\right)  =0,\ \ \ \mathrm{tr}%
\left(  \Upsilon_{d}^{(k)}\Upsilon_{d}^{(m)}\right)  =2\delta_{km},\nonumber
\end{align}
a basis of $\mathcal{L}_{d}$ consisting of the identity operator
$\mathbb{I}_{d}$ on $\mathcal{H}_{d}$ and a tuple
\begin{equation}
\Upsilon_{d}:=\left(  \Upsilon_{d}^{(1)},...,\Upsilon_{d}^{(d^{2}-1)}\right)
\end{equation}
of mutually orthogonal traceless Hermitian operators in $\mathcal{L}_{d}.$
Examples of operator bases $\{\mathbb{I}_{d},\  \Upsilon_{d}\}$ where elements
are non Hermitian were introduced in \cite{14}.

For every qudit observable $W\in\mathcal{L}_{d},$ $W=W^{\dag},$ the
decomposition in a basis $\mathfrak{B}_{\Upsilon_{d}}$ is given by%
\begin{align}
W  &  =\mathrm{tr}\left(  W\right)  \frac{\mathbb{I}}{d}\text{\textrm{ }%
}+\text{ }p_{\Upsilon_{d}}\cdot \Upsilon_{d},\text{ \ \ }p_{\Upsilon_{d}}%
\cdot \Upsilon_{d}:=\sum_{j=1}^{d^{2}-1}p_{\Upsilon_{d}}^{(j)}\Upsilon
_{d}^{(j)},\label{2.1}\\
p_{\Upsilon_{d}}^{(k)}  &  =\frac{1}{2}\mathrm{tr}\left(  \Upsilon_{d}%
^{(k)}W\right)  \in\mathbb{R},\text{ \ \ \ }p_{\Upsilon_{d}}:=(p_{\Upsilon
_{d}}^{(1)},...,p_{\Upsilon_{d}}^{(d^{2}-1)})\in\mathbb{R}^{d^{2}-1},\nonumber
\end{align}
and has the form of the representation via vector $p_{\Upsilon_{d}}%
\in\mathbb{R}^{d^{2}-1}$, satisfying the relation
\begin{equation}
\mathrm{tr}\left(  W^{2}\right)  =\frac{1}{d}\left(  \mathrm{tr}W\right)
^{2}\text{ }+\text{ }2\left\Vert p_{\Upsilon_{d}}\right\Vert _{\mathbb{R}%
^{d^{2}-1}}^{2}. \label{2.2}%
\end{equation}
This implies that, for an observable $W,$ the norm of the vector
$p_{\Upsilon_{d}}$ in decomposition (\ref{2.1}) does not depend on which
operator basis $\mathfrak{B}_{\Upsilon_{d}}$ of type (\ref{2}) is used in this
decomposition:%
\begin{equation}
\left\Vert p_{\Upsilon_{d}}\right\Vert _{\mathbb{R}^{d^{2}-1}}^{2}%
=||p_{\Upsilon_{d}^{\prime}}||_{\mathbb{R}^{d^{2}-1}}^{2},\text{ \ \ }%
\forall \Upsilon_{d},\Upsilon_{d}^{\prime}. \label{2_2}%
\end{equation}

\begin{notation}
For the vector in $\mathbb{R}^{d^{2}-1}$\ with components $\mathrm{tr}%
(\Upsilon_{d}^{(j)}W),$ $j=1,...,(d^{2}-1),$ we further use notation
$\mathrm{tr}(\Upsilon_{d}W),$ for short.
\end{notation}

The most known decomposition via a basis of type (\ref{2}) is the generalized
Gell-Mann representation \cite{6,7,9,10,15,16,17} specified in (\ref{2.1}) by
the tuple
\begin{equation}
\Lambda_{d}:=(\Lambda_{d}^{(1)},...,\Lambda_{d}^{(d^{2}-1)}) \label{2_3}%
\end{equation}
of the generalized Gell-Mann operators $\Lambda_{d}^{(k)}$ on $\mathbb{C}^{d}$
which are the higher-dimensional extensions of the Pauli operators
$\sigma:=(\sigma_{1},\sigma_{2},\sigma_{3})$ on $\mathbb{C}^{2}$ and the
Gell-Mann operators on $\mathbb{C}^{3}.$ For the product of the generalized
Gell-Mann operators, the decomposition in the basis $\mathfrak{B}_{\Lambda
_{d}}$ is given by%
\begin{equation}
\Lambda_{d}^{(k)}\Lambda_{d}^{(m)}=2\mathbb{\delta}_{km}\frac{\mathbb{I}}%
{d}\text{ \ }\mathbb{+}\text{ \ }\sum_{l}\left(  g_{kml}^{(\Lambda_{d}%
)}+if_{kml}^{(\Lambda_{d})}\right)  \Lambda_{d}^{(l)},\text{\ \ }\forall k,m,
\label{3}%
\end{equation}
and implies
\begin{align}
\lbrack\Lambda_{d}^{(k)},\Lambda_{d}^{(m)}]  &  =2i\sum_{l}\text{ }%
f_{kml}^{(\Lambda_{d})}\Lambda_{d}^{(l)},\label{3---1}\\
\Lambda_{d}^{(k)}\circ\Lambda_{d}^{(m)}  &  :=\Lambda_{d}^{(k)}\circ
\Lambda_{d}^{(m)}+\Lambda_{d}^{(m)}\circ\Lambda_{d}^{(k)}=4\mathbb{\delta
}_{km}\frac{\mathbb{I}}{d}\text{ }+\text{ }2\sum_{l}g_{kml}^{(\Lambda_{d}%
)}\Lambda_{d}^{(l)},\nonumber
\end{align}
where the real constants%
\begin{equation}
g_{kml}^{(\Lambda_{d})}=\frac{1}{4}\mathrm{tr}\left\{  \left(  \Lambda
_{d}^{(k)}\circ\Lambda_{d}^{(m)}\right)  \Lambda_{d}^{(l)}\right\}  ,\text{
\ \ }f_{kml}^{(\Lambda_{d})}=\frac{1}{4i}\mathrm{tr}\left\{  \left[
\Lambda_{d}^{(k)},\Lambda_{d}^{(m)}\right]  \Lambda_{d}^{(l)}\right\}  ,
\label{3.3'}%
\end{equation}
are symmetric and antisymmetric, respectively, under the permutation of
indices and constitute \emph{the} \emph{structure constants of group}
\textrm{SU(d).}

For $d=2$ and the tuple $\Lambda_{2}\equiv\sigma=(\sigma_{1},\sigma_{2}%
,\sigma_{3})$ of the Pauli qubit operators, all symmetric constants
$g_{kml}^{(\sigma)}=0$ while the antisymmetric constants have the form
$f_{kml}^{(\sigma)}=\varepsilon_{kml}$ where $\varepsilon_{kml}:=(e_{k}%
,e_{m},e_{l})$ are the components of the Levi-Chivita symbol, defined via the
mixed product of the corresponding elements of the standard basis of
$\mathbb{R}^{3}.$

Except for the tuple $\Lambda_{d}$ of the generalized Gell-Mann operators on
$\mathbb{C}^{d}$, $d\geq2,$ possible tuples of operators specified in
(\ref{2}), for example, include: (i) for $\mathcal{H}_{2}=\mathbb{C}^{2}$, the
operator tuple $\left(  \sigma_{+},\sigma_{-},\sigma_{3}\right)  $ where
$\sigma_{\pm}=\frac{\sigma_{1}+\sigma_{2}}{\sqrt{2}}$ ; (ii) for
$\mathcal{H}_{d_{1}\times d_{2}}=\mathbb{C}^{d_{1}}\otimes\mathbb{C}^{d_{2}},$
$d_{1},d_{2}\geq2,$ the tuple $\Gamma_{\Lambda_{d_{1}}\otimes\Lambda_{d_{2}}}$
of operators%
\begin{equation}
\Lambda_{d_{1}}^{(j)}\otimes\frac{\mathbb{I}_{d_{2}}}{\sqrt{d_{2}}},\text{
\ \ }\frac{\mathbb{I}_{d_{1}}}{\sqrt{d_{1}}}\otimes\Lambda_{d_{2}}%
^{(k)},\text{ \ \ }\frac{1}{\sqrt{2}}\text{ }\Lambda_{d_{1}}^{(j)}%
\otimes\Lambda_{d_{2}}^{(k)},\text{ \ \ }j=1,...,d_{1},\text{ \ }%
k=1,...,d_{2},
\end{equation}
for which the renormalized version of decomposition (\ref{2.1}), namely:%
\begin{align}
W  &  =\mathrm{tr}(W)\frac{\mathbb{I}_{d_{1}}\otimes\mathbb{I}_{d_{2}}}%
{d_{1}d_{2}}\text{ }+\text{ }\sqrt{\frac{d_{1}-1}{2d_{1}d_{2}^{2}}}\text{
}\left(  r_{1}\cdot\Lambda_{d_{1}}\right)  \otimes\mathbb{I}_{d_{2}}\text{
}+\text{ }\sqrt{\frac{d_{2}-1}{2d_{1}^{2}d_{2}}}\text{ }\mathbb{I}_{d_{1}%
}\otimes\left(  r_{2}\cdot\Lambda_{d_{2}}\right) \\
&  +\text{ }\sqrt{\frac{\left(  d_{1}-1\right)  \left(  d_{2}-1\right)
}{4d_{1}d_{2}}}\sum_{i,j}\mathcal{T}_{ij}\text{ }\Lambda_{d_{1}}%
^{(i)}\mathbb{\otimes}\Lambda_{d_{2}}^{(j)},\nonumber
\end{align}
constitutes a generalization to higher dimensions of the Pauli representation
in the two-qubit case.

For each tuple $\Upsilon_{d}$ $\neq\Lambda_{d}$ of mutually orthogonal
traceless Hermitian operators on $\mathbb{C}^{d}$, all its elements
$\Upsilon_{d}^{(k)}$ admit the generalized Gell-Mann representation
\begin{equation}
\Upsilon_{d}^{(k)}=\mathrm{v}_{\Lambda_{d}}^{(k)}\cdot\Lambda_{d},\text{
\ \ }\mathrm{v}_{\Lambda_{d}}^{(k)}=\frac{1}{2}\mathrm{tr}(\Lambda_{d}%
\Upsilon_{d}^{(k)})\in\mathbb{R}^{d^{2}-1}, \label{5"}%
\end{equation}
where $\left\{  \mathrm{v}_{\Lambda_{d}}^{(k)}\in\mathbb{R}^{d^{2}%
-1},\ k=1,...,(d^{2}-1)\right\}  $ is an orthonormal basis of $\mathbb{R}%
^{d^{2}-1}$, different from its standard basis. Similarly to (\ref{3}) the
decomposition in basis $\mathfrak{B}_{\Upsilon_{d}}$ of the product
$\Upsilon_{d}^{(k)}\Upsilon_{d}^{(m)}$ reads%
\begin{align}
\Upsilon_{d}^{(k)}\Upsilon_{d}^{(m)}  &  =\frac{2}{d}\mathbb{\delta}%
_{km}\mathbb{I}_{d}\text{ \ }\mathbb{+}\text{ \ }\sum_{l}\left(
g_{kml}^{(\Upsilon_{d})}\text{ }+\text{ }if_{kml}^{(\Upsilon_{d})}\right)
\Upsilon_{d}^{(l)},\text{\ \ \ }\forall k,m,\label{6}\\
g_{kml}^{(\Upsilon_{d})}  &  =\frac{1}{4}\mathrm{tr}\left\{  \left(
\Upsilon_{d}^{(k)}\circ \Upsilon_{d}^{(m)}\right)  \Upsilon_{d}^{(l)}\right\}
,\text{ \ \ }f_{kml}^{(\Upsilon_{d})}=\frac{1}{4i}\mathrm{tr}\left\{  \left[
\Upsilon_{d}^{(k)},\Upsilon_{d}^{(m)}\right]  \text{ }\Upsilon_{d}%
^{(l)}\right\}  ,\nonumber
\end{align}
where the real constants $g_{kml}^{(\Upsilon_{d})},$ $f_{kml}^{(\Upsilon_{d}%
)}$ are symmetric and antisymmetric with respect to the permutation of indices
and are expressed via the symmetric and antisymmetric structure constants
$g_{j_{1}j_{2}j_{3}}^{(\Lambda_{d})}$and $f_{j_{1}j_{2}j_{3}}^{(\Lambda_{d})}$
of group SU(d), given in (\ref{3}), via the relations
\begin{align}
g_{kml}^{(\Upsilon_{d})}  &  =\sum_{j_{1},j_{2},j_{3}}\left(  \mathrm{v}%
_{\Lambda_{d}}^{(k)}\right)  _{j_{1}}\left(  \mathrm{v}_{\Lambda_{d}}%
^{(m)}\right)  _{j_{2}}\left(  \mathrm{v}_{\Lambda_{d}}^{(l)}\right)  _{j_{3}%
}\text{ }g_{j_{1}j_{2}j_{3}}^{(\Lambda_{d})},\label{6.4}\\
\text{\ }f_{kml}^{(\Upsilon_{d})}  &  =\sum_{j_{1},j_{2},j_{3}}\left(
\mathrm{v}_{\Lambda_{d}}^{(k)}\right)  _{j_{1}}\left(  \mathrm{v}_{\Lambda
_{d}}^{(m)}\right)  _{j_{2}}\left(  \mathrm{v}_{\Lambda_{d}}^{(l)}\right)
_{j_{3}}\text{ }f_{j_{1}j_{2}j_{3}}^{(\Lambda_{d})}.\nonumber
\end{align}

For representation (\ref{2.1}) of qudit observables $W$ and $\widetilde{W}$ on
$\mathcal{H}_{d}$, specified for operator tuples $\Upsilon_{d}$ and
$\Upsilon_{d}^{\prime}$ in (\ref{2}):
\begin{align}
W  &  =\mathrm{tr}\left(  W\right)  \frac{\mathbb{I}}{d}\text{\textrm{ }%
}+\text{ }p_{\Upsilon_{d}}\cdot \Upsilon_{d}=\mathrm{tr}\left(  W\right)
\frac{\mathbb{I}}{d}\text{ }+\text{ }p_{\Upsilon_{d}^{\prime}}\cdot
\Upsilon_{d}^{\prime},\label{7}\\
\widetilde{W}  &  =\mathrm{tr}(\widetilde{W})\frac{\mathbb{I}}{d}\text{
}+\text{ }\widetilde{p}_{\Upsilon_{d}}\cdot \Upsilon_{d}=\mathrm{tr(}%
\widetilde{W})\frac{\mathbb{I}}{d}\text{ }+\text{ }\widetilde{p}_{\Upsilon
_{d}^{\prime}}\cdot \Upsilon_{d}^{\prime},\nonumber
\end{align}
relation (\ref{2_2}) implies
\begin{equation}
\left\Vert p_{\Upsilon_{d}}\right\Vert _{\mathbb{R}^{d^{2}-1}}^{2}=\left\Vert
p_{\Upsilon_{d}^{\prime}}\right\Vert _{\mathbb{R}^{d^{2}-1}}^{2},\text{
\ \ \ }||\widetilde{p}_{\Upsilon_{d}}||_{\mathbb{R}^{d^{2}-1}}^{2}%
=||\widetilde{p}_{\Upsilon_{d}^{\prime}}||_{\mathbb{R}^{d^{2}-1}}^{2}.
\label{xx}%
\end{equation}
Moreover, since, similarly to (\ref{5"}),
\begin{equation}
\Upsilon_{d}^{(k)}=\mathrm{v}_{\Upsilon_{d}^{\prime}}^{(k)}\cdot \Upsilon
_{d}^{\prime},\text{ \ \ }\mathrm{v}_{\Upsilon_{d}^{\prime}}^{(k)}=\frac{1}%
{2}\mathrm{tr}\left(  \Upsilon_{d}^{\prime}\Upsilon_{d}^{(k)}\right)
\in\mathbb{R}^{d^{2}-1}, \label{7"}%
\end{equation}
where $\left\{  \mathrm{v}_{\Upsilon_{d}^{\prime}}^{(k)}\in\mathbb{R}%
^{d^{2}-1},\ k=1,...,(d^{2}-1)\right\}  $ is an orthonormal basis of
$\mathbb{R}^{d^{2}-1},$ from (\ref{7"}) it follows%
\begin{align}
p_{\Upsilon_{d}^{\prime}}^{(j)}  &  =\sum_{k=1}^{d^{2}-1}\left(
\mathrm{v}_{\Upsilon_{d}^{\prime}}^{(k)}\right)  _{j}\text{ }p_{\Upsilon_{d}%
}^{(k)},\text{ \ \ \ }\widetilde{p}_{\Upsilon_{d}^{\prime}}^{(j)}=\sum
_{k=1}^{d^{2}-1}\left(  \mathrm{v}_{\Upsilon_{d}^{\prime}}^{(k)}\right)
_{j}\text{ }\widetilde{p}_{\Upsilon_{d}}^{(k)},\label{8}\\
p_{\Upsilon_{d}^{\prime}}\cdot\widetilde{p}_{\Upsilon_{d}^{\prime}}  &
=p_{\Upsilon_{d}}\cdot\widetilde{p}_{\Upsilon_{d}},\nonumber
\end{align}
where $\left[  T_{jk}\right]  :=[(\mathrm{v}_{\Upsilon_{d}^{\prime}}%
^{(k)})_{j}]$ is an orthogonal matrix.

Denote by $\mathfrak{O}_{\omega}\subset\mathcal{L}_{d}$ the set of all qudit
observables on $\mathcal{H}_{d}$ with a fixed value $\omega=\mathrm{tr}%
(W)\in\mathbb{R}$ of trace. Since representation (\ref{2.1}) is a
decomposition via a basis $\mathfrak{B}_{\Upsilon_{d}}$, and, for all
observables $W\in\mathfrak{O}_{\omega},$ the decomposition coefficient at
element $\mathbb{I}_{d}\in\mathfrak{B}_{\Upsilon_{d}}$ is fixed, the mapping
\begin{equation}
W\mapsto p_{\Upsilon_{d}}=\frac{1}{2}\mathrm{tr}\left(  \Upsilon_{d}W\right)
\in\mathbb{R}^{d^{2}-1},\text{ \ \ }W\in\mathfrak{O}_{\omega}, \label{9}%
\end{equation}
due to (\ref{2.1}) is injective and, for all $d\geq2$ and any tuple
$\Upsilon_{d}$ of operators satisfying relations in (\ref{2}), establishes the
one-to-one correspondence
\begin{equation}
\mathfrak{O}_{\omega}\leftrightarrow\mathfrak{I}_{\mathfrak{O}_{\omega}%
}^{(\Upsilon_{d})} \label{9.1}%
\end{equation}
between set $\mathfrak{O}_{\omega}$ and its image $\mathfrak{I}_{\mathfrak{O}%
_{\omega}}^{(\Upsilon_{d})}\subset\mathbb{R}^{d^{2}-1}$ under the injective
mapping (\ref{9}).

Relations (\ref{7})--(\ref{9.1}) imply.

\begin{proposition}
Let $\mathfrak{O}_{\omega}\subset\mathcal{L}_{d}$ be the set of qudit
observables with a fixed value $\omega=\mathrm{tr}(W)$ of trace. Under
representations (\ref{2.1}) specified for arbitrary tuples $\Upsilon_{d}%
\neq \Upsilon_{d}^{\prime}$ of operators, satisfying relations in (\ref{2}),
the images $\mathfrak{I}_{\mathfrak{O}_{\omega}}^{(\Upsilon_{d})}%
\subset\mathbb{R}^{d^{2}-1}$ and $\mathfrak{I}_{\mathfrak{O}_{\omega}%
}^{(\Upsilon_{d}^{\prime})}\subset$ $\mathbb{R}^{d^{2}-1}$ of set
$\mathfrak{O}_{\omega}$ are isometrically isomorphic.
\end{proposition}

\section{Bloch vectors}

Let $\mathcal{X}_{d}\subset\mathcal{L}_{d}$ be the set of all traceless qudit
observables $X$ on $\mathcal{H}_{d}$ with eigenvalues in $[-1,1]$ and
$\mathfrak{S}_{d}\subset\mathcal{L}_{d}$ be the set of all qudit states
(density operators) $\rho_{d}$ on $\mathcal{H}_{d},$ that is, positive
Hermitian operators with the unit trace:
\begin{equation}
\rho_{d}=\rho_{d}^{\dagger},\text{ \ \ }\rho_{d}\geq0,\text{ \ \ }%
\mathrm{tr}\left(  \rho_{d}\right)  =1.
\end{equation}

For qudit observables $X\in\mathcal{X}_{d},$ representation (\ref{2.1})
reduces to
\begin{equation}
X=x_{\Upsilon_{d}}\cdot \Upsilon_{d},\text{ \ \ \ \ }x_{\Upsilon_{d}}=\frac
{1}{2}\mathrm{tr}\left(  \Upsilon_{d}X\right)  \in\mathbb{R}^{d^{2}-1}.
\label{10}%
\end{equation}
Replacing in (\ref{10}) $x_{\Upsilon_{d}}\rightarrow\sqrt{\frac{d}{2}%
}n_{\Upsilon_{d}}$, we rewrite this representation in the form \cite{16}%
\begin{equation}
X=\sqrt{\frac{d}{2}}\left(  n_{\Upsilon_{d}}\cdot \Upsilon_{d}\right)
,\ \ \ n_{\Upsilon_{d}}=\sqrt{\frac{1}{2d}}\text{ }\mathrm{tr}\left(
\Upsilon_{d}X\right)  \in\mathbb{R}^{d^{2}-1}, \label{12}%
\end{equation}
which implies%
\begin{equation}
\mathrm{tr}\left(  X^{2}\right)  =d\left\Vert n_{\Upsilon_{d}}\right\Vert
_{\mathbb{R}^{d^{2}-1}}^{2}. \label{13}%
\end{equation}

For qudit states $\rho_{d}\in\mathfrak{S}_{d},$ representation (\ref{2.1})
takes the form
\begin{equation}
\rho_{d}=\frac{\mathbb{I}_{d}}{d}\text{ \ }\mathbb{+}\text{ \ }p_{\Upsilon
_{d}}\cdot \Upsilon_{d},\text{ \ \ }p_{\Upsilon_{d}}=\frac{1}{2}\mathrm{tr}%
\left(  \Upsilon_{d}\rho_{d}\right)  \in\mathbb{R}^{d^{2}-1}, \label{11}%
\end{equation}
and the renormalization $p_{\Upsilon_{d}}\rightarrow\sqrt{\frac{d-1}{2d}%
}r_{\Upsilon_{d}}$\ leads to the representation
\begin{align}
\rho_{d}  &  =\frac{\mathbb{I}_{d}}{d}\text{ \ }\mathbb{+}\text{ \ }%
\sqrt{\frac{d-1}{2d}}\left(  r_{\Upsilon_{d}}\cdot \Upsilon_{d}\right)
,\label{14}\\
r_{\Upsilon_{d}}  &  =\sqrt{\frac{d}{2(d-1)}}\text{ }\mathrm{tr}\left(
\rho_{d}\Upsilon_{d}\right)  \in\mathbb{R}^{d^{2}-1},\nonumber
\end{align}
for which%
\begin{equation}
\mathrm{tr}\left(  \rho_{d}^{2}\right)  =\frac{1}{d}+\frac{d-1}{d}\left\Vert
r_{\Upsilon_{d}}\right\Vert _{\mathbb{R}^{d^{2}-1}}^{2}. \label{15}%
\end{equation}

For the qubit case $(d=2)$ and the operator basis $\{\mathbb{I}_{2},\sigma\}$
comprised of the Pauli operators on $\mathbb{C}^{2}$, representations
(\ref{12}) and (\ref{14}) reduce to the well-known Bloch representations for
qubit states and traceless qubit observables \cite{1,2,3}:%
\begin{align}
\rho_{2}  &  =\frac{\mathbb{I}_{2}\text{ }\mathbb{+}\text{ }p_{\sigma}%
\cdot\sigma}{2},\text{ \ \ }p_{\sigma}=\mathrm{tr}\left(  \rho_{2}%
\sigma\right)  \in\mathbb{R}^{3},\label{16}\\
X  &  =n_{\sigma}\cdot\sigma,\text{ \ \ }n_{\sigma}=\frac{1}{2}\mathrm{tr}%
\left(  \sigma X\right)  \in\mathbb{R}^{3}.\nonumber
\end{align}
For a unit vector $\left\Vert n\right\Vert _{\mathbb{R}^{3}}=1,$ the traceless
qubit observable $n\cdot\sigma:=\sigma_{n}$ has eigenvalue $\pm1$ and is
interpreted as a projection $\sigma_{n}$ of a qubit spin along a direction
$n\in\mathbb{R}^{3}$.

\begin{notation}
In the Bloch representation (\ref{16}), vector $p_{\sigma}\in\mathbb{R}^{3}$
is called the Bloch vector (coherence vector) for a qubit state. For an
arbitrary $d\geq2$ and an arbitrary operator tuple $\Upsilon_{d}$, for
definiteness, we also further refer to vectors $n_{\Upsilon_{d}}%
,r_{\Upsilon_{d}}\in R^{d^{2}-1}$ in representations (\ref{12}) and (\ref{14}
) as (generalized) Bloch vectors and, if it is clear from a context, we
\ omit, for short, subscript $\Upsilon_{d}$ in their notation.
\end{notation}

Due to this terminology, in Proposition 1 the image $\mathfrak{I}%
_{\mathcal{X}_{d}}^{(\Upsilon_{d})}$ of the set $\mathcal{X}_{d}$ under
representation (\ref{12}) constitutes the set of Bloch vectors for all qudit
observables $X\in\mathcal{X}_{d}$ under this representation while the image
$\mathfrak{I}_{\mathfrak{S}_{d}}^{(\Upsilon_{d})}$ of the set $\mathfrak{S}%
_{d}$ under representation (\ref{14}) -- the set of the Bloch vectors for all
qudit states $\rho_{d}\in\mathfrak{S}_{d}$.

Since $\mathrm{tr}\left(  X^{2}\right)  \leq d,$ $\forall X\in\mathcal{X}%
_{d},$ and $\mathrm{tr}\left(  \rho_{d}^{2}\right)  \leq1$, $\forall\rho
_{d}\in\mathfrak{S}_{d},$ relations (\ref{13}), (\ref{15}) imply.

\begin{proposition}
[Necessary conditions]Let $\Upsilon_{d}$ be a tuple of qudit operators
satisfying relations in (\ref{2}). For each traceless qudit observable $X$
with eigenvalues in $[-1,1],$ the Bloch vector $n_{\Upsilon_{d}}\in
\mathbb{R}^{d^{2}-1},$ $d\geq2,$ in representation (\ref{12}) is necessarily%
\begin{equation}
\left\Vert n\right\Vert _{\mathbb{R}^{d^{2}-1}}\leq1, \label{18}%
\end{equation}
where the equality holds only for the Bloch vectors corresponding by
(\ref{12}) to traceless qudit observables with all its eigenvalues equal to
$\pm1.$ For any qudit state $\rho_{d}$, the Bloch vector $r_{\Upsilon_{d}}%
\in\mathbb{R}^{d^{2}-1}$ in representation (\ref{14}) is
necessarily\footnote{This condition is a generalization of the necessary
condition \cite{6,7} for the Bloch vector of a qudit state under the
generalized Gell-Mann representation.}%
\begin{equation}
\left\Vert r\right\Vert _{\mathbb{R}^{d^{2}-1}}\leq1, \label{19}%
\end{equation}
where the equality holds only for the Bloch vectors, corresponding by
(\ref{14}) to pure qudit states.
\end{proposition}

The necessary conditions (\ref{18}) and (\ref{19}) mean that, for all
$d\geq2,$ set $\mathfrak{J}_{\mathcal{X}_{d}}^{(\Upsilon_{d})}$ of the Bloch
vectors for all observables $X\in\mathcal{X}_{d}$ under representation
(\ref{12}) and set $\mathfrak{I}_{\mathfrak{S}_{d}}^{(\Upsilon_{d})}$ of the
Bloch vectors for all qudit states under representation (\ref{14}) constitute
subsets of the unit ball:%
\begin{align}
\mathfrak{J}_{\mathcal{X}_{d}}^{(\Upsilon_{d})}  &  \subseteq\{n\in
\mathbb{R}^{d^{2}-1}\mid\left\Vert n\right\Vert _{\mathbb{R}^{d^{2}-1}}%
\leq1\},\label{20}\\
\mathfrak{I}_{\mathfrak{S}_{d}}^{(\Upsilon_{d})}  &  \subseteq\{r\in
\mathbb{R}^{d^{2}-1}\mid\left\Vert r\right\Vert _{\mathbb{R}^{d^{2}-1}}%
\leq1\}.\nonumber
\end{align}

Let $r,\widetilde{r}\in\mathfrak{I}_{\mathfrak{S}_{d}}^{(\Upsilon_{d})}$ be
the Bloch vectors of qudit states $\rho_{d},\widetilde{\rho}_{d}%
\in\mathfrak{S}_{d}$ under representation (\ref{14}). Taking into account that
$\mathrm{tr}(\Upsilon_{d}^{(k)}\Upsilon_{d}^{(m)})=2\delta_{km}$, we have
$\mathrm{tr}\left(  \rho_{d}\widetilde{\rho}_{d}\right)  =\frac{1}{d}%
+\frac{d-1}{d}\left(  r\cdot\widetilde{r}\right)  \geq0.$ Therefore,
$(r\cdot\widetilde{r})\geq-\frac{1}{d-1}$.

Similarly, let observables $X,$ $\widetilde{X}\in\mathcal{X}_{d}$ be mutually
orthogonal in $\mathcal{L}_{d}$ and $n,\widetilde{n}\in\mathfrak{I}%
_{\mathcal{X}_{d}}^{(\Upsilon_{d})}$ be their Bloch vectors under
representation (\ref{12}). Then by (\ref{12}) and the mutual orthogonality
$\mathrm{tr}(X\widetilde{X})=d\left(  n\cdot\widetilde{n}\right)  =0.$

This implies the following properties, characterizing the Bloch vectors sets
$\mathfrak{I}_{\mathfrak{S}_{d}}^{(\Upsilon_{d})}$and $\mathfrak{I}%
_{\mathcal{X}_{d}}^{(\Upsilon_{d})}$.

\begin{proposition}
Let $\Upsilon_{d}$ be an arbitrary tuple of traceless Hermitian operators on
$\mathcal{H}_{d}$ satisfying relations in (\ref{2}). If vectors
$r,\widetilde{r}\in\mathfrak{I}_{\mathfrak{S}_{d}}^{(\Upsilon_{d})}%
\subset\mathbb{R}^{d^{2}-1}$ are the Bloch vectors of qudit states $\rho
_{d},\widetilde{\rho}_{d}\in\mathfrak{S}_{d}$ in representation (\ref{14}),
then their scalar product
\begin{equation}
r\cdot\widetilde{r}\geq-\frac{1}{d-1}. \label{22}%
\end{equation}
If vectors $n,\widetilde{n}\in\mathfrak{I}_{\mathcal{X}_{d}}^{(\Upsilon_{d}%
)}\subset\mathbb{R}^{d^{2}-1}$ are the Bloch vectors in representation
(\ref{12}) of observables $X,$ $\widetilde{X}\in\mathcal{X}_{d}$, mutually
orthogonal in space $\mathcal{L}_{d},$ then their scalar product
\begin{equation}
n\cdot\widetilde{n}=0. \label{22_1}%
\end{equation}

\end{proposition}

For the Bloch vectors of qudit states under the generalized Gell-Mann
representation, condition (\ref{22}) was presented in \cite{5,6,7}.

The expectation of a quantum observable $X$ in a qudit state $\rho_{d}%
\in\mathfrak{S}_{d}$ has the form
\begin{equation}
\mathrm{Ex}_{\rho_{d}}(X):=\mathrm{tr}\left(  \rho_{d}X\right)  . \label{22.1}%
\end{equation}
Substituting into the right hand-side of (\ref{22.1}) representations
(\ref{12}) and (\ref{14}) for an observable $X\in\mathcal{X}_{d}$ and a state
$\rho_{d}\in\mathfrak{S}_{d}$, respectively, we come to the following
representation%
\begin{equation}
\mathrm{Ex}_{\rho_{d}}(X)=\sqrt{d-1}\left(  r\cdot n\right)  \label{x}%
\end{equation}
of the quantum expectation (\ref{22.1}) via the scalar product of the Bloch
vectors $n\in\mathfrak{J}_{\mathcal{X}_{d}}^{(\Upsilon_{d})},$ $r\in
\mathfrak{J}_{\mathfrak{S}_{d}}^{(\Upsilon_{d})}$ bijectively corresponding to
an observable $X\in\mathcal{X}_{d}$ and a state $\rho_{d}\in\mathfrak{S}_{d}$
under representations (\ref{12}) and (\ref{14}), respectively.

\begin{remark}
Expressed in terms of Bloch vectors, the quantum analogs of bipartite Bell
inequalities for correlation functions constitute linear combinations of
scalar products of the corresponding Bloch vectors, for details, see our
results in \cite{16,18}.
\end{remark}

Note that, for an observable $X\in\mathcal{X}_{d}$, expectation $\left\vert
\mathrm{tr}\left(  \rho_{d}X\right)  \right\vert \leq1.$ This and relation
(\ref{x}) imply.

\begin{proposition}
Let $n\in\mathfrak{J}_{\mathcal{X}_{d}}^{(\Upsilon_{d})},$ $r\in
\mathfrak{J}_{\mathfrak{S}_{d}}^{(\Upsilon_{d})}$ be the Bloch vectors of an
observable $X\in\mathcal{X}_{d}$ and a state $\rho_{d}\in\mathfrak{S}_{d}$
under representations (\ref{12}) and (\ref{14}), respectively. Then
\begin{equation}
\sqrt{d-1}\left(  r\cdot n\right)  \leq1. \label{22.3}%
\end{equation}

\end{proposition}

From relation (\ref{22.3}) and Proposition 2 it follows that, for $d>2,$ an
observable $X\in\mathcal{X}_{d}$ with eigenvalues $\pm1$ and a pure state
$\rho_{d}\in\mathfrak{S}_{d}$ cannot be described by the same unit vector in
representations (\ref{12}) and (\ref{14}).

\begin{proposition}
Let $\mathfrak{J}_{\mathcal{X}_{d}}^{(\Upsilon_{d})}$ be the set of Bloch
vectors for all qudit observables $X\in\mathcal{X}_{d}$ under representation
(\ref{12}) and $\mathfrak{J}_{\mathfrak{S}_{d}}^{(\Upsilon_{d})}$ be the set
of the Bloch vectors for all qudit states $\rho_{d}\in\mathfrak{S}_{d}$ under
representation (\ref{14}). Then for all $d>2$
\begin{equation}
\left\{  n\in\mathfrak{J}_{\mathcal{X}_{d}}^{(\Upsilon_{d})}\mid\left\Vert
n\right\Vert _{\mathbb{R}^{d^{2}-1}}=1\right\}  \cap\left\{  r\in
\mathfrak{J}_{\mathfrak{S}_{d}}^{(\Upsilon_{d})}\mid\left\Vert r\right\Vert
_{\mathbb{R}^{d^{2}-1}}=1\right\}  =\varnothing.
\end{equation}

\end{proposition}

For our further consideration, we need to generalize the statement of Lemma 1
in \cite{16}, formulated for the tuple $\Lambda_{d}$ of the generalized
Gell-Mann operators on $\mathbb{C}^{d}$, to the case of an arbitrary operator
tuple $\Upsilon_{d}$ in (\ref{2}). As we stress in \cite{16}, the proofs of
the main statements in Section 2 of this article do not involve the specific
forms of the generalized Gell-Mann operators and hold for every operator tuple
$\Upsilon_{d}$ with elements satisfying the relations:
\begin{equation}
\Upsilon_{d}^{(k)}=\left(  \Upsilon_{d}^{(k)}\right)  ^{\dagger}%
\neq0,\ \ \mathrm{tr}\left(  \Upsilon_{d}^{(k)}\right)  =0,\ \ \mathrm{tr}%
(\Upsilon_{d}^{(k)}\Upsilon_{d}^{(m)})=2\delta_{km},\ \ \ k,m=1,...,(d^{2}-1),
\label{25}%
\end{equation}
that is, for any operator tuple $\Upsilon_{d}$ specified in (\ref{2}).

Denote by
\begin{equation}
\left\Vert X\right\Vert _{0}:=\sup_{\left\Vert \psi\right\Vert _{\mathcal{H}%
_{d}}=1}\left\Vert X\psi\right\Vert _{\mathcal{H}_{d}}=\max_{\lambda_{m}%
(X)}\left\vert \lambda_{m}(X)\right\vert \label{25.1}%
\end{equation}
the operator norm of a qudit observable $X$ with eigenvalues $\lambda_{m}(X)$.
The following statement is a generalization of Lemma 1 in \cite{16}.

\begin{proposition}
For each tuple $\Upsilon_{d}=(\Upsilon_{d}^{(1)},...,\Upsilon_{d}^{(d^{2}%
-1)})$ of traceless Hermitian operators on $\mathcal{H}_{d}$ satisfying
relations (\ref{25}), the upper and the lower bounds on the operator norm
\begin{equation}
\sqrt{\frac{2}{d}}\left\Vert p\right\Vert _{\mathbb{R}^{d^{2}-1}}%
\leq\left\Vert p\cdot \Upsilon_{d}\right\Vert _{0}\text{ }\leq\text{ }%
\sqrt{\frac{2(d-1)}{d}}\left\Vert p\right\Vert _{\mathbb{R}^{d^{2}-1}}
\label{26}%
\end{equation}
of a traceless Hermitian qudit operator $\left(  p\cdot \Upsilon_{d}\right)  $
hold for all vectors $p\in\mathbb{R}^{d^{2}-1}$ and all dimensions $d\geq2.$
\end{proposition}

Bounds (\ref{26}) imply.

\begin{corollary}
For each tuple $\Upsilon_{d}=(\Upsilon_{d}^{(1)},...,\Upsilon_{d}^{(d^{2}%
-1)})$ of traceless Hermitian operators on $\mathcal{H}_{d}$ satisfying
relations (\ref{25}) and all $d\geq2:$\newline\textrm{(a)} $\left\Vert
p\right\Vert _{\mathbb{R}^{d^{2}-1}}\leq\sqrt{\frac{1}{d-1}}$ $\ \ \Rightarrow
$ $\ \ \left\Vert p\cdot \Upsilon_{d}\right\Vert _{0}\leq\sqrt{\frac{2}{d}};$
\newline\textrm{(b)}$\left\Vert p\right\Vert _{\mathbb{R}^{d^{2}-1}}\leq
\frac{1}{d-1}$ $\ \ \Rightarrow$ $\ \ \left\Vert p\cdot \Upsilon_{d}\right\Vert
_{0}\leq\sqrt{\frac{2}{d(d-1)}};$\newline\textrm{(c)} $\left\Vert p\right\Vert
_{\mathbb{R}^{d^{2}-1}}\leq1$ $\ \ \Rightarrow$ $\ \ \left\Vert p\cdot
\Upsilon_{d}\right\Vert _{0}\leq\sqrt{\frac{2(d-1)}{d}};$\newline\textrm{(d)}
$\left\Vert p\cdot \Upsilon_{d}\right\Vert _{0}\leq\sqrt{\frac{2}{d}}$
$\ \ \Rightarrow$ \ $\ \left\Vert p\right\Vert _{\mathbb{R}^{d^{2}-1}}\leq
1;$\newline\textrm{(e)} $\left\Vert p\cdot \Upsilon_{d}\right\Vert _{0}%
\leq\sqrt{\frac{2}{d}}\left\Vert p\right\Vert _{\mathbb{R}^{d^{2}-1}}$
$\ \ \Rightarrow$ $\ \left\Vert p\cdot \Upsilon_{d}\right\Vert _{0}=\sqrt
{\frac{2}{d}}\left\Vert p\right\Vert _{\mathbb{R}^{d^{2}-1}}.$
\end{corollary}

\section{Bloch vectors of traceless qudit observables}

Under the generalized Gell-Mann representation $X=\sqrt{\frac{d}{2}}\left(
n\cdot\Lambda_{d}\right)  ,$ we specified the set $\mathfrak{J}_{\mathcal{X}%
_{d}}^{(\Lambda_{d})}$ of Bloch vectors for all qudit observables
$X\in\mathcal{X}_{d}$ in article \cite{16}. This representation is a
particular case of representation (\ref{12}) with $\Upsilon_{d}\rightarrow
\Lambda_{d}$. As we stressed above, the proofs of our main statements in
Section 2 of \cite{16}, formulated for the generalized Gell-Mann
representation $X=\sqrt{\frac{d}{2}}\left(  n\cdot\Lambda_{d}\right)  $, hold
for representation (\ref{12}) with every operator tuple $\Upsilon_{d}=\left(
\Upsilon_{1},...,\Upsilon_{d^{2}-1}\right)  $ of operators on $\mathcal{H}%
_{d}$ satisfying relations (\ref{25}).

For decomposition (\ref{12}) via any basis $\mathfrak{B}_{\Upsilon_{d}}$ of
type (\ref{2}), the generalization of Theorem 1 in \cite{16} and the above
Propositions 1--3 and Corollary 2 imply.

\begin{theorem}
Let $\Upsilon_{d}=\left(  \Upsilon_{d}^{(1)},...,\Upsilon_{d}^{(d^{2}%
-1)}\right)  ,$ $d\geq2,$ be a tuple of traceless Hermitian operators on
$\mathcal{H}_{d}$ satisfying relations (\ref{25}) and $\mathcal{X}_{d}$ be the
set of all traceless qudit observables with the operator norm $\left\Vert
X\right\Vert _{0}\leq1.$ The representation
\begin{equation}
X=\sqrt{\frac{d}{2}}\text{ }(n\cdot \Upsilon_{d}) \label{27}%
\end{equation}
establishes the one-to-one correspondence\ $\mathcal{X}_{d}\leftrightarrow
$\ $\mathfrak{J}_{\mathcal{X}_{d}}^{(\Upsilon_{d})}$ between observables
$X\in\mathcal{X}_{d}$ and vectors $n\in\mathbb{R}^{d^{2}-1}$ in the set
\begin{equation}
\mathfrak{J}_{\mathcal{X}_{d}}^{(\Upsilon_{d})}=\left\{  n\in\mathbb{R}%
^{d^{2}-1}\mid\left\Vert n\cdot \Upsilon\right\Vert _{0}\leq\sqrt{\frac{2}{d}%
}\right\}  , \label{28}%
\end{equation}
which is a subset of the unit ball:
\begin{equation}
\mathfrak{J}_{\mathcal{X}_{d}}^{(\Upsilon_{d})}\subseteq\left\{
n\in\mathbb{R}^{d^{2}-1}\mid\left\Vert n\right\Vert _{\mathbb{R}^{d^{2}-1}%
}\leq1\right\}  , \label{28__}%
\end{equation}
and contains the ball of radius $\frac{1}{\sqrt{d-1}}:$%
\begin{equation}
\mathfrak{J}_{\mathcal{X}_{d}}^{(\Upsilon_{d})}\supseteq\left\{
n\in\mathbb{R}^{d^{2}-1}\mid\left\Vert n\right\Vert _{\mathbb{R}^{d^{2}-1}%
}\leq\frac{1}{\sqrt{d-1}}\right\}  \label{28.1}%
\end{equation}
The boundary of $\mathfrak{J}_{\mathcal{X}_{d}}^{(\Upsilon_{d})}$ has the form%
\begin{equation}
\partial\mathfrak{J}_{\mathcal{X}_{d}}^{(\Upsilon_{d})}=\left\{
n\in\mathbb{R}^{d^{2}-1}\mid\left\Vert n\cdot \Upsilon\right\Vert _{0}%
=\sqrt{\frac{2}{d}}\right\}  . \label{28.2}%
\end{equation}
For $d\geq3,$ the geometry of the set%
\begin{equation}
\mathfrak{J}_{\mathcal{X}_{d}}^{(\Upsilon_{d})}\cap\left\{  n\in
\mathbb{R}^{d^{2}-1}\mid\frac{1}{\sqrt{d-1}}<\left\Vert n\right\Vert
_{\mathbb{R}^{d^{2}-1}}\leq1\right\}
\end{equation}
is rather complicated. The maximal norm of a vector $n\in\mathfrak{J}%
_{\mathcal{X}_{d}}^{(\Upsilon_{d})}$ is equal to $1$ if a dimension $d\geq2$
is even and to $\sqrt{\frac{d-1}{d}}$ if a dimension $d>2$ is odd. Under the
one-to-one correspondence $\mathcal{X}_{d}\leftrightarrow\mathfrak{J}%
_{\mathcal{X}_{d}}^{(\Upsilon_{d})}$, established by representation
(\ref{27}), the sets%
\begin{equation}
\left\{  \ X\in\mathcal{X}_{d}\mid\lambda_{m}(X)=\pm1,\text{ \ }%
m=1,...,d\right\}  \leftrightarrow\left\{  \ n\in\mathfrak{J}_{\mathcal{X}%
_{d}}^{(\Upsilon_{d})}\mid\left\Vert n\right\Vert =1\right\}  \label{30}%
\end{equation}
and are not empty if and only if a dimension $d\geq2$ is even.
\end{theorem}

From Theorem 1 it follows that the boundary of $\mathfrak{J}_{\mathcal{X}_{d}%
}^{(\Upsilon_{d})}$ contains unit vectors%
\begin{equation}
\partial\mathfrak{J}_{\mathcal{X}_{d}}^{(\Upsilon_{d})}\cap\left\{
n\in\mathbb{R}^{d^{2}-1}\mid\left\Vert n\right\Vert _{\mathbb{R}^{d^{2}-1}%
}=1\right\}  \neq\varnothing\label{32}%
\end{equation}
if and only if a qudit dimension $d\geq2$ is even.

\section{Bloch vectors of qudit states}

Under representation (\ref{14}) specified with the operator tuple $\Lambda
_{d}$ of the generalized Gell-Mann operators (i. e. under the generalized
Gell-Mann representation), the set $\mathfrak{I}_{\mathfrak{S}_{d}}%
^{(\Lambda_{d})}$ of Bloch vectors set $\mathfrak{I}_{\mathfrak{S}_{d}%
}^{(\Lambda_{d})}$ was specified in \cite{6,7} where it was proved that a
Hermitian operator $\tau_{d}=\frac{\mathbb{I}}{d}$ $\mathbb{+}$ $\sqrt
{\frac{d-1}{2d}}\left(  r_{\Lambda_{d}}\cdot\Lambda_{d}\right)  \ $with the
unit trace $\mathrm{tr}\left(  \tau_{d}\right)  =1$ is positive $\tau_{d}%
\geq0$, hence, constitutes a qudit state, if an only if
\begin{equation}
a_{j}(\tau_{d})\geq0,\text{ \ }\ j=2,...,d, \label{34}%
\end{equation}
where coefficients $a_{j}(\tau_{d})$ are derived in \cite{6,7} via the
recursive relations and have the forms:
\begin{align}
2!a_{2}  &  =1-\mathrm{tr}\left(  \tau_{d}^{2}\right)  ,\label{33}\\
3!a_{3}  &  =1-3\mathrm{tr}\left(  \tau_{d}^{2}\right)  +2\mathrm{tr}\left(
\tau_{d}^{3}\right)  ,\nonumber\\
4!a_{4}  &  =1-6\mathrm{tr}\left(  \tau_{d}^{2}\right)  +8\mathrm{tr}\left(
\tau_{d}^{3}\right)  +3\left(  \mathrm{tr}\left(  \tau_{d}^{2}\right)
\right)  ^{2}-6\mathrm{tr}\left(  \tau_{d}^{4}\right)  ,\nonumber\\
5!a^{5}  &  =\cdots.\nonumber
\end{align}
Moreover, the operator $\tau_{d}$ constitutes a pure qudit state if and only
if $a_{j}(\tau_{d})=0,$ for all $j=2,...,d.$

The proofs of these results in \cite{6,7} do not involve the specific forms of
the generalized Gell-Mann operators but are only based on relations (\ref{25})
and the application of Newton's formulas for sums of the powers of roots
$\lambda_{j}=1,...,d$ of the characteristic equation for the matrix
representation of operator $\tau_{d}$. This means that relations (\ref{34}),
(\ref{33}) are also true for the decomposition%
\begin{equation}
\tau_{d}=\frac{\mathbb{I}}{d}\text{ \ }\mathbb{+}\text{ \ }\sqrt{\frac
{d-1}{2d}}\left(  r_{\Upsilon_{d}}\cdot \Upsilon_{d}\right)  ,\text{ \ \ }%
r\in\mathbb{R}^{d^{2}-1}, \label{32_}%
\end{equation}
where $\Upsilon_{d}$ is an arbitrary tuple of operators satisfying conditions
(\ref{25}). Substituting (\ref{32_}) into (\ref{33}) and taking into account
relations (\ref{3}) we derive:
\begin{align}
2!a_{2}^{(\Upsilon_{d})}(r)  &  =\frac{d-1}{d}(1-\left\Vert r\right\Vert
_{\mathbb{R}^{d^{2}-1}}^{2}),\label{36}\\
3!a_{3}^{(\Upsilon_{d})}(r)  &  =\frac{(d-1)(d-2)}{d^{2}}\left(  1-3\left\Vert
r\right\Vert _{\mathbb{R}^{d^{2}-1}}^{2}\right)  +2\frac{d-1}{d}\sqrt
{\frac{d-1}{2d}}\sum_{i,j,k}g_{ijk}^{(\Upsilon_{d})}r_{i}r_{j}r_{k}%
,\nonumber\\
4!a_{4}^{(\Upsilon_{d})}(r)  &  =\frac{(d-1)(d-2)(d-3)}{d^{3}}(1-6\left\Vert
r\right\Vert _{\mathbb{R}^{d^{2}-1}}^{2})+3\frac{(d-1)^{2}(d-2)}{d^{3}%
}\left\Vert r\right\Vert _{\mathbb{R}^{d^{2}-1}}^{4}\nonumber\\
&  +\frac{8(d-1)(d-3)}{d^{2}}\sqrt{\frac{d-1}{2d}}\sum_{i,j,k}g_{ijk}%
^{(\Upsilon_{d})}r_{i}r_{j}r_{k}\nonumber\\
&  -3\frac{\left(  d-1\right)  ^{2}}{d^{2}}\sum_{i,j,l,m,k}g_{kij}%
^{(\Upsilon_{d})}g_{klm}^{(\Upsilon_{d})}r_{i}r_{j}r_{k}r_{l},\nonumber\\
5!a_{5}^{(\Upsilon_{d})}(r)  &  =\cdots,\nonumber
\end{align}
where, for short of notations, we omit the lower index $\Upsilon_{d}$ at
$r_{\Upsilon_{d}}.$ Note that by (\ref{2_2}) the norm $\left\Vert r\right\Vert
_{\mathbb{R}^{d^{2}-1}}^{2}$of the Bloch vector of a state $\tau_{d}$ is the
same for representation (\ref{14}) via different operator tuples $\Upsilon
_{d}$.

From (\ref{34}) and (\ref{36}) it follows that, under representation
(\ref{14}),\ set $\mathfrak{I}_{\mathfrak{S}_{d}}^{(\Upsilon_{d})}$ of the
Bloch vectors of all qudit states and its subset $\mathfrak{I}_{\mathfrak{S}%
_{d}^{pure}}^{(\Upsilon_{d})}\subset\mathfrak{I}_{\mathfrak{S}_{d}}%
^{(\Upsilon_{d})}$ of the Bloch vectors of all pure qudit states are given by
\begin{align}
\mathfrak{I}_{\mathfrak{S}_{d}}^{(\Upsilon_{d})}  &  =\left\{  r\in
\mathbb{R}^{d^{2}-1}\mid a_{j}^{(\Upsilon_{d})}(r)\geq0,\text{ \ \ }%
j=2,...,d\right\}  ,\label{37}\\
\mathfrak{I}_{\mathfrak{S}_{d}^{pure}}^{(\Upsilon_{d})}  &  =\left\{
r\in\mathbb{R}^{d^{2}-1}\mid a_{j}^{(\Upsilon_{d})}(r)=0,\text{ \ \ }%
j=2,...,d\right\}  , \label{38}%
\end{align}
respectively,

For the qubit case ($d=2)$ and the operator type $\Upsilon_{2}$ $=\Lambda
_{2}\equiv\sigma=(\sigma_{1},\sigma_{2},\sigma_{3}),$ the sets (\ref{37}) and
(\ref{38}) reduce correspondingly, to the unit ball and the unit sphere in
$\mathbb{R}^{3}$ -- the well-known results from the Bloch vectors formalism
for qubit states$.$ For higher dimensions, the geometrical properties of set
$\mathfrak{I}_{\mathfrak{S}_{3}}^{(\Lambda_{3})}$ of the Bloch vectors of all
qutrit states under the generalized Gell-Mann representation, also, the
two-dimensional and three-dimensional sections of set $\mathfrak{I}%
_{\mathfrak{S}_{d}}^{(\Lambda_{d})}$ for $d=3,4$ were analyzed in
\cite{51,6,11,12,13,15}.

However, the specification of sets $\mathfrak{I}_{\mathfrak{S}_{d}}%
^{(\Upsilon_{d})}$ and $\mathfrak{I}_{\mathfrak{S}_{d}^{pure}}^{(\Upsilon
_{d})}$ via the systems of algebraic equations in (\ref{37}) and (\ref{38})
does not allow to characterize these sets in a compact unified\ analytical
form for all $d\geq2$, also, to find their general geometry properties.

\emph{In what follows, we introduce for set }$\mathfrak{I}_{\mathfrak{S}_{d}%
}^{(\Upsilon_{d})}$ \emph{of the Bloch vectors for all qudit states} \emph{and
its subset }$\mathfrak{I}_{\mathfrak{S}_{d}^{pure}}^{(\Upsilon_{d})}%
\subset\mathfrak{I}_{\mathfrak{S}_{d}}^{(\Upsilon_{d})}$ \emph{of the Bloch
vectors for all pure qudit states,} \emph{the new compact expressions in terms
of operator norms.}

\emph{These new expressions have} \emph{the unified forms for all }$d\geq2$
\emph{and reveal the general geometry properties of sets }$\mathfrak{I}%
_{\mathfrak{S}_{d}}^{(\Upsilon_{d})}$ \emph{and} $\mathfrak{I}_{\mathfrak{S}%
_{d}^{pure}}^{(\Upsilon_{d})}$ \emph{which for }$d\geq3$ \emph{cannot be
analytically extracted from the systems of algebraic equations specified in
(\ref{37}), (\ref{38}).}

Denote by $\lambda_{m}^{(+)}(r)>0$ and $\lambda_{m}^{(-)}(r)\leq0$ the
positive and non-positive eigenvalues of a traceless Hermitian operator
$\left(  r\cdot \Upsilon_{d}\right)  $ and by $k_{\lambda_{m}}$ -- the
multiplicity of an eigenvalue $\lambda_{m}.$ The spectral decomposition of a
Hermitian operator (\ref{32_}) reads%
\begin{equation}
\tau_{d}=\sum_{\lambda_{m}^{(+)}}\left(  \frac{1}{d}\text{ }\mathbb{+}\text{
}\sqrt{\frac{d-1}{2d}}\lambda_{m}^{(+)}(r)\right)  \mathrm{E}_{\lambda
_{m}^{(+)}}\text{ }+\text{ }\sum_{\lambda_{m}^{(-)}}\left(  \frac{1}{d}\text{
}\mathbb{-}\text{ }\sqrt{\frac{d-1}{2d}}\left\vert \lambda_{m}^{(-)}%
(r)\right\vert \right)  \mathrm{E}_{\lambda_{m}^{(-)}}, \label{39}%
\end{equation}
where $\mathrm{E}_{\lambda_{m}}$ is the spectral projection of a Hermitian
operator $(r\cdot \Upsilon_{d})$ corresponding to its eigenvalue $\lambda
_{m}(r)$. Relation (\ref{39}) implies that, for the Hermitian operator
(\ref{32_}), all its eigenvalues are given by
\begin{align}
\xi_{m}(r)  &  =\left(  \frac{1}{d}\text{ \ }\mathbb{+}\text{ \ }\sqrt
{\frac{d-1}{2d}}\lambda_{m}^{(+)}(r)\right)  >0,\text{ }\label{39.1}\\
\eta_{j}(r)  &  =\left(  \frac{1}{d}\text{ \ }\mathbb{-}\text{ \ }\sqrt
{\frac{d-1}{2d}}\left\vert \lambda_{j}^{(-)}(r)\right\vert \right)  ,\text{
}\nonumber
\end{align}
and have multiplicities $k_{\lambda_{m}}$ of the corresponding eigenvalues
$\lambda_{m}^{(\pm)}(r)$. Therefore, a Hermitian operator $\tau_{d}$
constitutes a qudit state if and only if all its eigenvalues $\eta_{j}(r)$ are
non-negative. From (\ref{39.1}) it follows that this is true if and only if%

\begin{equation}
\max_{\lambda_{j}^{(-)}}\left\vert \lambda_{j}^{(-)}(r)\right\vert \leq
\sqrt{\frac{2}{d(d-1)}}. \label{40}%
\end{equation}

Recall \cite{4} that any qudit observable $Z$ admits the decomposition
\begin{align}
Z  &  =Z^{(+)}-Z^{(-)},\text{ \ \ }Z^{(+)},Z^{(-)}\geq0,\label{41}\\
Z^{(+)}Z^{(-)}  &  =Z^{(-)}Z^{(+)}=0,\nonumber
\end{align}
via positive Hermitian operators $Z^{(\pm)}\geq0.$ This, in particular, refers
to traceless qudit observables $\left(  r\cdot \Upsilon\right)  $ with the
operator norm $\left\Vert r\cdot \Upsilon_{d}\right\Vert _{0}$ satisfying
bounds (\ref{25.1}).

In view of this and relation (\ref{25.1}), the necessary and sufficient
condition (\ref{40}) is equivalent to $\left\Vert \left(  r\cdot \Upsilon
_{d}\right)  ^{(-)}\right\Vert _{0}\leq\sqrt{\frac{2}{d(d-1)}}$ in expression
(\ref{32_}). The latter, in turn, implies $\mathrm{tr}\left(  \tau_{d}%
^{2}\right)  \leq1,$ and, hence,%
\begin{equation}
\mathrm{tr}\left(  \tau_{d}^{2}\right)  =\frac{1}{d}+\frac{d-1}{d}\left\Vert
r\right\Vert _{\mathbb{R}^{d^{2}-1}}^{2}\leq1\text{ \ }\Leftrightarrow\text{
\ }\left\Vert r\right\Vert _{\mathbb{R}^{d^{2}-1}}^{2}\leq1, \label{42}%
\end{equation}
so that by item (c) of Corollary 1%
\begin{equation}
\left\Vert r\cdot \Upsilon_{d}\right\Vert _{0}\leq\sqrt{\frac{2(d-1)}{d}}.
\label{43}%
\end{equation}

Furthermore, a Hermitian operator (\ref{32_}) is a pure qudit state if and
only if it is positive and $\mathrm{tr}\left(  \tau_{d}^{2}\right)
=1\Leftrightarrow\left\Vert r\right\Vert _{\mathbb{R}^{d^{2}-1}}^{2}=1.$
Moreover, in this case $\tau_{d}$ has only the eigenvalue equal to $1$ with
multiplicity $1$ and the eigenvalue equal to $0$ with multiplicity $(d-1)$. In
notation of (\ref{39.1}) these are $\xi=1$ with multiplicity $k_{\xi}=1$\ and
$\eta=0$ with multiplicity $k_{\eta}=d-1.$ But the latter is possible iff
$\left\Vert \left(  r\cdot \Upsilon\right)  ^{(-)}\right\Vert _{0}=\sqrt
{\frac{2}{d(d-1)}}.$

Relations (\ref{39})--(\ref{43}) prove the following statement.

\begin{proposition}
A Hermitian operator (\ref{32_}) constitutes a qudit state if and only if%
\begin{equation}
\left\Vert \left(  r\cdot \Upsilon_{d}\right)  ^{(-)}\right\Vert _{0}\leq
\sqrt{\frac{2}{d(d-1)}} \label{46}%
\end{equation}
and this condition implies
\begin{equation}
\left\Vert r\right\Vert _{\mathbb{R}^{d^{2}-1}}^{2}\leq1. \label{47}%
\end{equation}
A Hermitian operator (\ref{32_}) constitutes a pure qudit state if and only
if
\begin{align}
\left\Vert \left(  r\cdot \Upsilon_{d}\right)  ^{(-)}\right\Vert _{0}  &
=\sqrt{\frac{2}{d(d-1)}},\label{48}\\
\left\Vert r\right\Vert _{\mathbb{R}^{d^{2}-1}}^{2}  &  =1.\nonumber
\end{align}

\end{proposition}

Since $\left\Vert \left(  r\cdot \Upsilon_{d}\right)  ^{(-)}\right\Vert
_{0}\leq\left\Vert r\cdot \Upsilon_{d}\right\Vert _{0},$ from (\ref{46}) and
item (b) of Corollary 1 it follows.

\begin{corollary}
[(Sufficient condition)]For all
\begin{equation}
\left\Vert r\right\Vert _{\mathbb{R}^{d^{2}-1}}^{2}\leq\frac{1}{d-1},
\label{49}%
\end{equation}
condition (\ref{46}) is fulfilled.
\end{corollary}

Propositions 2, 3, 7 and Corollary 2 imply.

\begin{theorem}
Let $\Upsilon_{d}=\left(  \Upsilon_{d}^{(1)},...,\Upsilon_{d}^{(d^{2}%
-1)}\right)  ,$ $d\geq2,$ be a tuple of traceless Hermitian operators on
$\mathcal{H}_{d}$ satisfying relations (\ref{25}). The representation
\begin{equation}
\rho_{d}=\frac{\mathbb{I}_{d}}{d}\text{ }\mathbb{+}\text{ }\sqrt{\frac
{d-1}{2d}}\left(  r\cdot \Upsilon_{d}\right)  \label{50}%
\end{equation}
establishes the one-to-one correspondence\ $\mathfrak{S}_{d}\leftrightarrow
$\ $\mathfrak{J}_{\mathfrak{S}_{d}}^{(\Upsilon_{d})}$ between qudit states
$\rho_{d}\in\mathfrak{S}_{d}$ and vectors $r\in\mathbb{R}^{d^{2}-1}$ in the
set%
\begin{equation}
\mathfrak{J}_{\mathfrak{S}_{d}}^{(\Upsilon_{d})}=\left\{  r\in\mathbb{R}%
^{d^{2}-1}\mid\left\Vert \left(  r\cdot \Upsilon_{d}\right)  ^{(-)}\right\Vert
_{0}\leq\sqrt{\frac{2}{d(d-1)}}\right\}  , \label{51}%
\end{equation}
which is a subset%
\begin{equation}
\mathfrak{J}_{\mathfrak{S}_{d}}^{(\Upsilon_{d})}\subseteq\left\{
r\in\mathbb{R}^{d^{2}-1}\mid\left\Vert r\right\Vert _{\mathbb{R}^{d^{2}-1}%
}\leq1\right\}
\end{equation}
of the unit ball and contains the ball of radius $\frac{1}{d-1}:$%
\begin{equation}
\mathfrak{J}_{\mathfrak{S}_{d}}^{(\Upsilon_{d})}\supseteq\left\{
r\in\mathbb{R}^{d^{2}-1}\mid\left\Vert r\right\Vert _{\mathbb{R}^{d^{2}-1}%
}\leq\frac{1}{d-1}\right\}  . \label{52}%
\end{equation}
For any two vectors $r_{1},r_{2}\in\mathfrak{J}_{\mathfrak{S}_{d}}%
^{(\Upsilon_{d})},$%
\begin{equation}
r_{1}\cdot r_{2}\geq-\frac{1}{d-1}.
\end{equation}
The boundary of $\mathfrak{J}_{\mathfrak{S}_{d}}^{(\Upsilon_{d})}$ has the
form
\begin{equation}
\partial\mathfrak{J}_{\mathfrak{S}_{d}}^{(\Upsilon_{d})}=\left\{
r\in\mathbb{R}^{d^{2}-1}\mid\left\Vert \left(  r\cdot \Upsilon_{d}\right)
^{(-)}\right\Vert _{0}=\sqrt{\frac{2}{d(d-1)}}\right\}  .
\end{equation}
For $d\geq3,$ the geometry of the set
\begin{equation}
\mathfrak{J}_{\mathfrak{S}_{d}}^{(\Upsilon_{d})}\cap\left\{  r\in
\mathbb{R}^{d^{2}-1}\mid\frac{1}{d-1}<\left\Vert r\right\Vert _{\mathbb{R}%
^{d^{2}-1}}\leq1\right\}
\end{equation}
is rather complicated. Under the one-to-one correspondence $\mathfrak{S}%
_{d}\leftrightarrow\mathfrak{J}_{\mathfrak{S}_{d}}^{(\Upsilon)}$, established
by representation (\ref{50}), pure qudit states are bijectively mapped to
vectors $r\in\mathbb{R}^{d^{2}-1}$ in the subset
\begin{equation}
\mathfrak{J}_{\mathfrak{S}_{d}^{pure}}^{(\Upsilon_{d})}=\left\{
r\in\mathbb{R}^{d^{2}-1}\mid\left\Vert \left(  r\cdot \Upsilon\right)
^{(-)}\right\Vert _{0}=\sqrt{\frac{2}{d(d-1)}},\text{ \ \ }\left\Vert
r\right\Vert _{\mathbb{R}^{d^{2}-1}}^{2}=1\right\}  \label{55}%
\end{equation}
of the unit sphere in $\mathbb{R}^{d^{2}-1}.\mathfrak{\ }$
\end{theorem}

\section{Evolution in time}

For the Bloch vector $r(t)\in\mathfrak{J}_{\mathfrak{S}_{d}}^{(\Upsilon)}$ of
a qudit state $\rho_{d}(t)$ under representation (\ref{14}), let us now
consider its evolution in time if a qudit system is isolated and if it is open.

Recall that if a qudit system is isolated, then, under a qudit Hamiltonian
$H_{d}(t)=H_{d}^{\dag}(t),$ the evolution of its state $\rho_{d}(t),$
$t>t_{0},$ in time is described by the relation%
\begin{equation}
\rho_{d}(t)=U(t,t_{0})\rho_{d}(t_{0})U^{\dagger}(t,t_{0}), \label{55_}%
\end{equation}
where $\rho_{d}(t_{0})$ is an initial state of a qudit system and $U(t,t_{0})$
is the unitary operator on $\mathcal{H}_{d}$, satisfying the Cauchy problem
for the Schr\"{o}dinger equation:%
\begin{align}
i\frac{d}{dt}U(t,t_{0})  &  =H_{d}(t)U(t,t_{0}),\text{ \ \ }t>t_{0}%
,\label{56}\\
U(t_{0},t_{0})  &  =\mathbb{I}_{d}.\nonumber
\end{align}
Eqs. (\ref{56}) and (\ref{55_}) imply that, for an isolated qudit system, the
time evolution of its state $\rho_{d}(t)$ is described by the solution of the
Liouville--von Neumann equation%
\begin{equation}
\frac{d}{dt}\rho_{d}(t)=-i\text{ }[H_{d}(t),\rho_{d}(t)],\text{ \ \ }t>t_{0},
\label{57"}%
\end{equation}
satisfying the initial condition $\rho_{d}(t_{0}).$

If, however, a qudit system is open, i. e. interacts with an environment,
then, in the Markovian case, the evolution in time of its state $\rho_{d}(t)$
is described by the Lindblad master equation \cite{19,4} which we take in the
following generalized form%
\begin{align}
\frac{d}{dt}\rho_{d}(t)  &  =-i\left[  \widetilde{H}_{d}(t),\rho
_{d}(t)\right]  +\sum_{k}\gamma_{k}\left(  L_{k}(t)\rho_{d}(t)L_{k}^{\dag
}(t)-\frac{1}{2}L_{k}^{\dag}(t)L_{k}(t)\circ\rho_{d}(t)\right)  ,\label{57.1}%
\\
\gamma_{k}  &  \geq0.\nonumber
\end{align}
Here, in the right hand side: (i) the first term describes the time evolution
under a qudit Hamiltonian $\widetilde{H}_{d}(t)=$ $\widetilde{H}_{d}^{\dagger
}(t),$ including, in general, a "bare" Hamiltonian $H_{d}(t)$ of a qudit
system and an additive due to its interaction with an environment; (ii) the
second term describes the dissipative part with, in general, nonstationary
operators $L_{k}(t)\in\mathcal{L}_{d}$; (iii) notation $A_{1}\circ A_{2}$ is
determined in (\ref{3}).

Taking into account that representation (\ref{14}) of a qudit state holds for
all moments of time:%
\begin{align}
\rho_{d}(t)  &  =\frac{\mathbb{I}_{d}}{d}\text{ }\mathbb{+}\text{ }\sqrt
{\frac{d-1}{2d}}\left(  r(t)\cdot \Upsilon_{d}\right)  ,\label{58}\\
r(t)  &  =\sqrt{\frac{d}{2(d-1)}}\text{ }\mathrm{tr}\left(  \rho
_{d}(t)\Upsilon_{d}\right)  \in\mathfrak{J}_{\mathfrak{S}_{d}}^{(\Upsilon
_{d})}\subset\mathbb{R}^{d^{2}-1},\text{ \ \ }t\geq t_{0},\nonumber
\end{align}
we have
\begin{equation}
\frac{d}{dt}r(t)=\sqrt{\frac{d}{2(d-1)}}\text{ }\mathrm{tr}\left(
\Upsilon_{d}\frac{d}{dt}\rho_{d}(t)\right)  . \label{58.2}%
\end{equation}

In what follows, based on relation (\ref{58.2}) and Eqs. (\ref{57"}) and
(\ref{57.1}), we specify the general equations describing the time evolution
of the Bloch vector $r(t)$ of a qudit system state $\rho_{d}(t)$ under
representation (\ref{58}) if a qudit system is isolated and if a qudit system
is open.

\subsection{Isolated qudit system}

For an isolated qudit system, the evolution in time of its state $\rho_{d}(t)$
under a Hamiltonian $H_{d}(t)$ is described by the Liouville--von Neumann
equation (\ref{57"}). This equation and relation (\ref{58.2}) imply%
\begin{equation}
\frac{d}{dt}r(t)=-i\sqrt{\frac{d}{2(d-1)}}\mathrm{tr}\left(  [H_{d}%
(t),\rho_{d}(t)]\text{ }\Upsilon_{d}\right)  . \label{59'}%
\end{equation}
Taking into account representation (\ref{58}) for state $\rho(t),$ the
decomposition (\ref{2.1}) of a general qudit Hamiltonian $H_{d}(t)$ in a basis
$\mathfrak{B}_{\Upsilon_{d}}:$%
\begin{align}
H_{d}(t)  &  =h_{0}(t)\frac{\mathbb{I}_{d}}{d}\text{ \ }\mathbb{+}\text{
\ }h(t)\cdot \Upsilon_{d}\text{ },\label{60}\\
h_{0}(t)  &  =\text{\textrm{tr}}\left(  H_{d}(t)\right)  \in\mathbb{R},\text{
\ \ }h(t)=\frac{1}{2}\text{\textrm{tr}}\left(  \Upsilon_{d}H_{d}(t)\right)
\in\mathbb{R}^{d^{2}-1},\nonumber
\end{align}
and relations (\ref{6}), we have%
\begin{align}
\sqrt{\frac{d}{2(d-1)}}\left[  H_{d}(t),\rho_{d}(t)\right]   &  =\frac{1}%
{2}\left[  \left(  h(t)\cdot \Upsilon_{d}\right)  ,\text{ }\left(
r(t)\cdot \Upsilon_{d}\right)  \right]  =\frac{1}{2}\sum_{k,m,l}h_{k}(t)\text{
}r_{m}(t)\left[  \Upsilon_{d}^{(k)},\Upsilon_{d}^{(m)}\right] \label{61}\\
&  =i\sum_{k,m,l}f_{kml}^{(\Upsilon_{d})}\text{ }h_{k}(t)r_{m}(t)\Upsilon
_{d}^{(l)},\nonumber
\end{align}
where constants $f_{kml}^{(\Upsilon_{d})}$ are defined in (\ref{6}).

The substitution of (\ref{61}) into relation (\ref{59'}) proves the following statement.

\begin{theorem}
Let $\Upsilon_{d}=\left(  \Upsilon_{d}^{(1)},...,\Upsilon_{d}^{(d^{2}%
-1)}\right)  ,$ $d\geq2,$ be a tuple of traceless Hermitian operators on
$\mathcal{H}_{d}$ satisfying conditions (\ref{25}) and Eq. (\ref{60}) be the
decomposition of a general nonstationary qudit Hamiltonian $H_{d}(t)$ in basis
$\mathfrak{B}_{\Upsilon_{d}}$.\ Under the time evolution of a qudit state
$\rho_{d}(t)$ due to the Liouville--von Neumann equation (\ref{57"}), the
evolution in time of its Bloch vector $r(t)\in\mathfrak{J}_{\mathfrak{S}_{d}%
}^{(\Upsilon_{d})}\subset\mathbb{R}^{d^{2}-1}$ in representation (\ref{58}) is
described by
\begin{align}
\frac{d}{dt}r(t)  &  =\mathbb{B}_{H_{d}}(t)r(t),\text{ \ \ }t>t_{0}%
,\label{62}\\
r(t_{0})  &  =\sqrt{\frac{d}{2(d-1)}}\mathrm{tr}\left(  \rho_{d}%
(t_{0})\Upsilon_{d}\right)  ,\nonumber
\end{align}
where $\mathbb{B}_{H_{d}}(t):\mathbb{R}^{d^{2}-1}\rightarrow\mathbb{R}%
^{d^{2}-1}$ is the the skew-symmetric linear operator defined via its matrix
representation in the standard basis of $\mathbb{R}^{d^{2}-1}:$
\begin{equation}
\mathbb{B}_{H_{d}}^{(lm)}(t)=-2\sum_{k}f_{lmk}^{(\Upsilon_{d})}\text{ }%
h_{k}(t)=-\mathbb{B}_{H_{d}}^{(ml)}(t), \label{62.1}%
\end{equation}
with constants $f_{lmk}^{(\Upsilon_{d})}$ given in (\ref{6}).
\end{theorem}

For the qubit case $(d=2)$ and the tuple $\Upsilon_{2}=\sigma,$ i.e. in case
of the Bloch representation (\ref{16}), Eq. (\ref{62}) reduces to
\begin{equation}
\frac{d}{dt}r(t)=2\text{ }h(t)\times r(t), \label{64}%
\end{equation}
where $h(t)\in\mathbb{R}^{3}$ is the vector in the decomposition
$H_{2}(t)=h_{0}(t)\frac{\mathbb{I}}{2}\ \mathbb{+}\ h(t)\cdot\sigma$ of a
general qubit Hamiltonian $H_{2}(t)$ in the operator basis $\{\mathbb{I}%
_{2},\sigma\}$ and notation $\left(  a\times c\right)  $ means the vector
product of vectors $a,c\in\mathbb{R}^{3}$.

Taking into account that by (\ref{64})
\begin{equation}
\frac{1}{2}\frac{d}{dt}\left\Vert r(t)\right\Vert _{\mathbb{R}^{d^{2}-1}}%
^{2}=r\cdot\left(  \mathbb{B}_{H_{d}}(t)r(t)\right)  ,\text{ \ \ }t\geq t_{0},
\end{equation}
and that operator $\mathbb{B}_{H_{d}}(t)$ is skew symmetric, we have $\frac
{d}{dt}\left\Vert r(t)\right\Vert _{\mathbb{R}^{d^{2}-1}}^{2}=0.$ This implies
the following statement.

\begin{proposition}
Under the unitary evolution (\ref{55_}), the norm of the Bloch vector
$r(t)\in\mathfrak{J}_{\mathfrak{S}_{d}}^{(\Upsilon_{d})}\subset\mathbb{R}%
^{d^{2}-1}$ of a qudit system state $\rho_{d}(t)$ in representation (\ref{58})
is invariant in time:%
\begin{equation}
\left\Vert r(t)\right\Vert _{\mathbb{R}^{d^{2}-1}}=\left\Vert r(t_{0}%
)\right\Vert _{\mathbb{R}^{d^{2}-1}},\text{ \ \ \ }t\geq t_{0}, \label{64.1}%
\end{equation}
for all $d\geq2.$
\end{proposition}

Note that by (\ref{xx}) the norm $\left\Vert r(t)\right\Vert _{\mathbb{R}%
^{d^{2}-1}}$ of the Bloch vector of a qudit state does not also depend on a
choice of an operator tuple $\Upsilon_{d}$ in representation (\ref{58})\ 

For the unitary operator $U(t,t_{0})$ in (\ref{56}), the differential
equations describing the evolution in time of its decomposition coefficients
under the generalized Gell-Mann representation were found by us recently in
\cite{16}, see there Eq. (26). The derivation in \cite{16} of these equations
does not involve the specific forms of the generalized Gell-Mann operators and
is only based only on the validity of relations (\ref{25}), which are,
however, true for all tuples $\Upsilon_{d}$ in operator bases of type (\ref{2}).

Therefore, according to our results in \cite{16}, for an arbitrary operator
tuple $\Upsilon_{d},$ the decomposition of $U(t,t_{0})$ in a
basis$\mathfrak{\ B}_{\Upsilon_{d}}$ takes the form
\begin{equation}
U(t,t_{0})=\exp\left\{  -i\int_{t_{0}}^{t}h_{0}(\tau)d\tau\right\}  \left(
u_{0}(t,t_{0})\mathbb{I}_{d}-i\sqrt{\frac{d}{2}}\ u(t,t_{0})\cdot \Upsilon
_{d}\right)  , \label{65}%
\end{equation}
where\footnote{Here, notation $u^{\prime}$ means the vector-column in
$\mathbb{C}^{d^{2}-1}$ comprised of components of vector $u=(u_{1}%
,...,u_{d^{2}-1})$ and $\overline{u}^{(j)}$ \ -- the complex conjugate of
$u_{j}\in\mathbb{C}.$}%
\begin{align}
\left\vert u_{0}(t,t_{0})\right\vert ^{2}+\left\Vert u^{\prime}(t,t_{0}%
)\right\Vert _{\mathbb{C}^{d^{2}-1}}^{2}  &  =1,\\
u_{0}(t,t_{0})\overline{u}^{(j)}(t,t_{0})-\overline{u}_{0}(t,t_{0}%
)u^{(j)}(t,t_{0})  &  =\sqrt{\frac{d}{2}}\sum_{k,m}\left(  f_{kmj}%
^{(\Upsilon_{d})}-ig_{kmj}^{(\Upsilon_{d})}\right)  u^{(k)}(t,t_{0}%
)\overline{u}^{(m)}(t,t_{0}),\nonumber\\
u_{0}(t,t_{0})  &  \in\mathbb{C},\ \ \ u_{j}(t,t_{0})\in\mathbb{C},\text{
\ \ }j=1,...,(d^{2}-1),\nonumber
\end{align}
and, under the time evolution of the unitary operator $U(t,t_{0})$ due to the
Schr\"{o}dinger equation (\ref{56}), the evolution in time of its
decomposition coefficients in representation (\ref{65})\ is described by the
system of linear ordinary differential equations:
\begin{align}
\overset{\cdot}{u}_{0}(t,t_{0})  &  =h(t)\cdot u(t,t_{0}),\label{66}\\
\frac{d}{dt}u_{j}(t,t_{0})  &  =-u_{0}(t,t_{0})h_{j}(t)+\sqrt{\frac{d}{2}}%
\sum_{m,k}\left(  f_{jkm}^{(\Upsilon_{d})}-ig_{jkm}^{(\Upsilon_{d})}\right)
h_{k}(t)u_{m}(t,t_{0}),\nonumber\\
u_{0}(t_{0},t_{0})  &  =1,\text{ \ \ }u_{j}(t_{0},t_{0})=0,\text{
\ \ }j=1,...,(d^{2}-1),\nonumber
\end{align}
where $g_{jkm}^{(\Upsilon_{d})}$ and $f_{jkm}^{(\Upsilon_{d})}$ are symmetric
and antisymmetric constants defined in (\ref{6}).

\subsection{Open qudit system}

Let a qudit system be open and the evolution in time of its state $\rho
_{d}(t)$ be described by the Lindblad master equation (\ref{57.1}). From
(\ref{58.2}) and (\ref{57.1}) it follows%
\begin{align}
\frac{d}{dt}r_{l}(t)  &  =\sqrt{\frac{d}{2(d-1)}}\text{ }\mathrm{tr}\left(
Z_{l}(t)\right)  ,\label{67}\\
Z_{l}(t)  &  =\Upsilon_{d}^{(l)}\left(  -i\left[  \widetilde{H}_{d}%
(t),\rho_{d}(t)\right]  +\sum_{k}\gamma_{k}\left(  L_{k}(t)\rho_{d}%
(t)L_{k}^{\dag}(t)-\frac{1}{2}L_{k}^{\dag}(t)L_{k}(t)\circ\rho_{d}(t)\right)
\right)  .\nonumber
\end{align}

Let $\widetilde{H}_{d}(t)=\widetilde{h}_{0}(t)\frac{\mathbb{I}}{d}$
\ $\mathbb{+}$ \ $\widetilde{h}(t)\cdot \Upsilon_{d}$ be the decomposition a
general Hamiltonian $\widetilde{H}_{d}(t)$ in a basis $\mathfrak{B}%
_{\Upsilon_{d}}$. Then, similarly to (\ref{61}), the commutator
\begin{equation}
\sqrt{\frac{d}{2(d-1)}}\left[  \widetilde{H}_{d}(t),\rho_{d}(t)\right]
=i\sum_{k,m,l}f_{kml}^{(\Upsilon_{d})}\widetilde{h}_{k}(t)r_{m}(t)\Upsilon
_{d}^{(l)}%
\end{equation}
and substituting this relation and decomposition (\ref{58}) into Eq.
(\ref{67}), we derive%
\begin{align}
\sqrt{\frac{d}{2(d-1)}}\text{ }\mathrm{tr}\left(  Z_{l}(t)\right)   &
=\sum_{m}\mathbb{B}_{\widetilde{H}_{d}}^{(lm)}(t)r_{m}(t)+\sqrt{\frac
{1}{2d(d-1)}}\sum_{k}\gamma_{k}\mathrm{tr}\left(  \Upsilon_{d}^{(l)}\left[
L_{k}(t),L_{k}^{\dag}(t)\right]  \right) \\
&  +\frac{1}{2}\sum_{k,m}\gamma_{k}r_{m}(t)\mathrm{tr}\left(  \Upsilon
_{d}^{(l)}\left(  L_{k}(t)\Upsilon_{d}^{(m)}L_{k}^{\dag}(t)-\frac{1}{2}%
L_{k}^{\dag}(t)L_{k}(t)\circ \Upsilon_{d}^{(m)}\right)  \right)  ,\nonumber
\end{align}
where matrix $\mathbb{B}_{\widetilde{H}_{d}}^{(lm)}(t)=-2\sum_{k}%
f_{lmk}^{(\Upsilon)}\widetilde{h}_{k}(t)$ is skew-symmetric. Noting further
that
\begin{align}
&  \mathrm{tr}\left(  L_{k}^{\dag}(t)\Upsilon_{d}^{(l)}L_{k}(t)\Upsilon
_{d}^{(m)}-\frac{1}{2}L_{k}^{\dag}(t)L_{k}(t)\left(  \Upsilon_{d}^{(l)}%
\circ \Upsilon_{d}^{(m)}\right)  \right) \label{70}\\
&  =\frac{1}{2}\mathrm{tr}\left(  \left[  \Upsilon_{d}^{(l)},L_{k}(t)\right]
\Upsilon_{d}^{(m)}L_{k}^{\dag}(t)+\left[  L_{k}^{\dag}(t),\Upsilon_{d}%
^{(l)}\right]  L_{k}(t)\Upsilon_{d}^{(m)}\right) \nonumber\\
&  =\operatorname{Re}\left(  \mathrm{tr}\left(  \left[  \Upsilon_{d}%
^{(l)},L_{k}(t)\right]  \Upsilon_{d}^{(m)}L_{k}^{\dag}(t)\right)  \right)
,\nonumber
\end{align}
we come to the following general statement.

\begin{theorem}
Let $\Upsilon_{d}=(\Upsilon_{d}^{(1)},...,\Upsilon_{d}^{(d^{2}-1)}),$
$d\geq2,$ be a tuple of traceless Hermitian operators on $\mathcal{H}_{d}$,
satisfying conditions (\ref{25}) and $\widetilde{H}_{d}(t)=\widetilde{h}%
_{0}(t)\frac{\mathbb{I}}{d}$ $\mathbb{+}$ $\widetilde{h}(t)\cdot \Upsilon_{d}$
be the decomposition of a general qudit Hamiltonian $\widetilde{H}_{d}(t)$ in
basis $\mathfrak{B}_{\Upsilon_{d}}$. Under the time evolution of a qudit state
$\rho_{d}(t)$ due to the Lindblad master equation (\ref{57.1}), the evolution
in time of its Bloch vector $r(t)\in\mathfrak{J}_{\mathfrak{S}_{d}}%
^{(\Upsilon_{d})}\subset\mathbb{R}^{d^{2}-1}$ in representation (\ref{58}) is
described by
\begin{align}
\frac{d}{dt}r(t)  &  =\left(  \mathbb{B}_{\widetilde{H}_{d}}(t)+\mathbb{B}%
_{dis}(t)\right)  r(t)+\sqrt{\frac{1}{2d(d-1)}}\sum_{k}\gamma_{k}%
\mathrm{tr}\left(  \text{ }\Upsilon_{d}\left[  L_{k}(t),L_{k}^{\dag
}(t)\right]  \right)  ,\label{71}\\
r(t_{0})  &  =\sqrt{\frac{d}{2(d-1)}}\mathrm{tr}\left(  \rho_{d}%
(t_{0})\Upsilon_{d}\right)  ,\nonumber
\end{align}
where $\mathbb{B}_{\widetilde{H}_{d}}(t):\mathbb{R}^{d^{2}-1}\rightarrow
\mathbb{R}^{d^{2}-1}$ and $\mathbb{B}_{dis}(t):\mathbb{R}^{d^{2}-1}%
\rightarrow\mathbb{R}^{d^{2}-1}$ are linear operators defined via their matrix
representations in the standard basis of $\mathbb{R}^{d^{2}-1}:$
\begin{align}
\mathbb{B}_{\widetilde{H}_{d}}^{(lm)}(t)  &  :=-2\sum_{k}f_{lmk}%
^{(\Upsilon_{d})}\widetilde{h}_{k}(t)=-\mathbb{B}_{\widetilde{H}_{d}}%
^{(ml)}(t),\label{71.1}\\
\mathbb{B}_{dis}^{(lm)}(t)  &  =\frac{1}{2}\sum_{k}\gamma_{k}\operatorname{Re}%
\left(  \mathrm{tr}\left(  [\Upsilon_{d}^{(l)},L_{k}(t)]\text{ }\Upsilon
_{d}^{(m)}L_{k}^{\dag}(t)\right)  \right)  ,\text{ \ \ }\gamma_{k}>0.
\label{71.2}%
\end{align}
Operator $\mathbb{B}_{\widetilde{H}_{d}}(t)$ is skew symmetric, constants
$f_{lmk}^{(\Upsilon_{d})}$ are given in (\ref{6}).
\end{theorem}

Let us analyze the evolution in time of the norm $\left\Vert r(t)\right\Vert
_{\mathbb{R}^{d^{2}-1}}$ of the Bloch vector of a qudit state $\rho_{d}(t).$
We multiply the left-hand and the right-hand sides of Eq. (\ref{71}) by
component $r_{l}(t),$ consider further the sum of the resulting expressions
over $l=1,....,(d^{2}-1),$ take into account $r(t)\cdot\left(  \mathbb{B}%
_{\widetilde{H}_{d}}(t)r(t)\right)  =0$\ since operator $\mathbb{B}%
_{\widetilde{H}_{d}}$ is skew symmetric and finally derive:
\begin{align}
\frac{1}{2}\frac{d}{dt}\left\Vert r(t)\right\Vert _{\mathbb{R}^{d^{2}-1}}^{2}
&  =\sqrt{\frac{1}{2d(d-1)}}\sum_{k}\gamma_{k}\mathrm{tr}\left(
X_{d}(t)\left[  L_{k}(t),L_{k}^{\dag}(t)\right]  \right) \label{73}\\
&  +\frac{1}{2}\sum_{k}\gamma_{k}\operatorname{Re}\left(  \mathrm{tr}\left(
\left[  X_{d}(t),L_{k}(t)\right]  \text{ }X_{d}(t)L_{k}^{\dag}(t)\right)
\right)  ,\nonumber\\
X_{d}(t)  &  :=r(t)\cdot \Upsilon_{d}.\nonumber
\end{align}
Since the qudit operator $X_{d}(t)$ is Hermitian at each moment time, its
spectral decomposition reads%
\begin{equation}
X_{d}=\sum_{j=1,...,d}x_{j}\text{ }|x_{j}\rangle\langle x_{j}|, \label{74}%
\end{equation}
where $\{|x_{j}\rangle\in\mathcal{H}_{d},$ $j=1,...,d\}$ is the orthonormal
basis comprised of the eigenvectors of operator $X_{d}$ corresponding to
eigenvalues $x_{1}\leq x_{2}\leq\cdots\leq x_{d}$. Here and further, for
short, we omit in Eqs. (\ref{75})--(\ref{78_1}) the dependence of time in
notations for operators $X_{d},$ $L_{k}$, eigenvalues $x_{j}$ and eigenvectors
$|x_{j}\rangle$.

Substituting (\ref{74}) into the right-hand side of (\ref{73}) we have%
\begin{align}
\left[  X_{d},L_{k}\right]   &  =\sum_{j=1,...,d}x_{j}\left\{  \text{ }%
|x_{j}\rangle\langle x_{j}|L_{k}-L_{k}|x_{j}\rangle\langle x_{j}|\right\}
,\label{75}\\
X_{d}L_{k}^{\dag}  &  =\sum_{i=1,...,d}x_{i}\text{ }|x_{i}\rangle\langle
x_{i}|L_{k}^{\dag},\nonumber\\
\mathrm{tr}\left(  \left[  X_{d},L_{k}\right]  X_{d}L_{k}^{\dag}\right)   &
=\sum_{j=1,...,d}\langle x_{j}|[X_{d},L_{k}]X_{d}L_{k}^{\dag}|x_{j}%
\rangle\nonumber\\
&  =\sum_{i,j}x_{j}x_{i}\text{ }\langle x_{j}|L_{k}|\text{ }x_{i}%
\rangle\langle x_{i}|L_{k}^{\dag}|x_{j}\rangle-\sum_{j,i}x_{i}^{2}\langle
x_{j}|L_{k}|x_{i}\rangle\langle x_{i}|L_{k}^{\dag}|x_{j}\rangle. \label{76}%
\end{align}
Taking into account that, in (\ref{76}),%
\begin{equation}
x_{j}x_{i}\leq\frac{1}{2}\left(  x_{j}^{2}+x_{i}^{2}\right)  ,\text{
\ \ \ }\sum_{n=1,...,d}|x_{n}\rangle\langle x_{n}|=\mathbb{I}_{d}, \label{77}%
\end{equation}
we derive
\begin{align}
\operatorname{Re}\left(  \mathrm{tr}\left(  [X_{d},L_{k}]X_{d}L_{k}^{\dag
}\right)  \right)   &  \leq\frac{1}{2}\sum_{i,j}\left(  x_{j}^{2}+x_{i}%
^{2}\text{ }\right)  \langle x_{j}|L_{k}|x_{i}\rangle\langle x_{i}|L_{k}%
^{\dag}|x_{j}\rangle-\sum_{j,i}x_{i}^{2}\langle x_{j}|L_{k}|x_{i}%
\rangle\langle x_{i}|L_{k}^{\dag}|x_{j}\rangle\label{78}\\
&  =\frac{1}{2}\sum_{j}x_{j}^{2}\text{ }\langle x_{j}|L_{k}L_{k}^{\dag}%
|x_{j}\rangle-\frac{1}{2}\sum_{i}x_{i}^{2}\langle x_{i}|L_{k}^{\dag}%
L|x_{i}\rangle\nonumber\\
&  =\frac{1}{2}\sum_{m}x_{m}^{2}\text{ }\langle x_{m}|[L_{k},L_{k}^{\dag
}]|x_{m}\rangle.\nonumber
\end{align}
Relations (\ref{73})--(\ref{78}) imply.

\begin{proposition}
Let, in the Lindblad equation (\ref{57.1}), each operator $L_{k}(t)$ be normal
$L_{k}(t)L_{k}^{\dag}(t)=L_{k}^{\dag}(t)L_{k}(t)$ at all moments $t>t_{0}$ of
time. Then, under the time evolution of a qudit state $\rho_{d}(t)$ due to the
Lindblad equation (\ref{57.1}), the norm $\left\Vert r(t)\right\Vert
_{\mathbb{R}^{d^{2}-1}}$ of its Bloch vector in representation (\ref{58}) is a
non-increasing function of time:%
\begin{equation}
\frac{d}{dt}\left\Vert r(t)\right\Vert _{\mathbb{R}^{d^{2}-1}}\leq0,\text{
\ \ }t>t_{0}.
\end{equation}

\end{proposition}

\section{Entanglement of a pure bipartite state}

Let $\rho_{d_{1}\times d_{2}}$ be a pure bipartite state on a Hilbert space
$\mathcal{H}_{d_{1}}\otimes\mathcal{H}_{d_{2}}$ with arbitrary dimensions
$\dim\mathcal{H}_{j}=d_{j}\geq2$ and
\begin{equation}
\rho_{d_{1}}:=\mathrm{tr}_{\mathcal{H}_{d_{2}}}\left(  \rho_{d_{1}\times
d_{2}}\right)  \text{, \ \ }\rho_{d_{2}}:=\mathrm{tr}_{\mathcal{H}_{1}}\left(
\rho_{d_{1}\times d_{2}}\right)  \label{01}%
\end{equation}
be the states on $\mathcal{H}_{d_{j}},$ $j=1,2,$ reduced from a state
$\rho_{d_{1}\times d_{2}}.$

By the Schmidt decomposition \cite{2,3,4}, the non-zero eigenvalues of the
reduced states $\rho_{d_{1}}$ and $\rho_{d_{2}}$ coincide, therefore,%
\begin{equation}
\mathrm{tr}\left(  \rho_{d_{1}}^{2}\right)  =\mathrm{tr}\left(  \rho_{d_{2}%
}^{2}\right)  . \label{02}%
\end{equation}
For a pure bipartite state $\rho_{d_{1}\times d_{2}},$ the parameter
\begin{equation}
\mathrm{C}_{\rho_{d_{1}\times d_{2}}}:=\sqrt{\alpha_{d_{1}\times d_{2}}\left(
1-\mathrm{tr}(\rho_{d_{j}}^{2})\right)  },\text{ \ \ }j=1,2,\text{
\ \ \ }\alpha_{d_{1}\times d_{2}}>0, \label{03}%
\end{equation}
where $\alpha_{d_{1}\times d_{2}}>0$ is some positive constant, is monotone
increasing in the entanglement of this state and constitutes an entanglement
measure called the concurrence \cite{20, 21}. \emph{A pure bipartite state
}$\rho_{d_{1}\times d_{2}}$\emph{ is separable iff }$C_{\rho_{d_{1}\times
d_{2}}}=0$\emph{ and entangled iff }$C_{\rho_{d_{1}\times d_{2}}}>0.$

In Eq. (\ref{03}), a choice of coefficient $\alpha_{d_{1}\times d_{2}}>0$
depends on a normalization of the concurrence $\mathrm{C}_{\rho_{d_{1}\times
d_{2}}}.$ In \cite{20}, coefficient $\alpha_{d_{1}\times d_{2}}$ is taken to
be equal to $2$ for all $d_{1},d_{2}\geq2$ \ -- similarly as it is for a
two-qubit pure state $\rho_{2\times2}$. However, below we choose the
normalization in (\ref{03}) such that, for a maximally\ entangled two-qudit
state $\rho_{d_{1}\times d_{2}}^{m\text{-}e},$ the concurrence\textrm{
C}$_{\rho_{d_{1}\times d_{2}}^{m\text{-}e}}=1,$ for all $d_{1},d_{2}\geq2.$ As
we prove below, the latter results in the coefficient $\alpha_{d_{1}\times
d_{2}}$ different from the value $2$ taken in \cite{20}.

For the reduced states $\rho_{d_{j}}$, consider their decompositions
(\ref{14})
\begin{equation}
\rho_{d_{j}}=\frac{\mathbb{I}_{d_{j}}}{d_{j}}\text{ }+\text{ }\sqrt
{\frac{d_{j}-1}{2d_{j}}}(r_{\rho_{d_{j}}}\cdot \Upsilon_{d_{j}}),\text{
\ \ \ }j=1,2, \label{05}%
\end{equation}
via arbitrary tuples $\Upsilon_{d_{j},}$,$j=1,2,$ of traceless Hermitian
operators satisfying relations (\ref{25}) and acting on $\mathcal{H}_{d_{1}}$
and $\mathcal{H}_{d_{2}},$ respectively. In decompositions (\ref{05}), the
Bloch vectors of the reduced state $\rho_{d_{j}}$ are given by
\[
r_{\rho_{d_{j}}}=\sqrt{\frac{d_{j}}{2(d_{j}-1)}}\text{ }\mathrm{tr}\left(
\rho_{d_{j}}\Upsilon_{d_{j}}\right)  \in\mathbb{R}^{d_{j}^{2}-1}%
\]
and by Theorem 2%
\begin{equation}
||r_{\rho_{d_{j}}}||_{\mathbb{R}^{d_{j}^{2}-1}}\leq1. \label{06}%
\end{equation}
The relation%
\begin{equation}
\mathrm{tr}(\rho_{d_{j}}^{2})=\frac{1}{d_{j}}+\frac{d_{j}-1}{d_{j}}%
||r_{\rho_{d_{j}}}||_{\mathbb{R}^{d_{j}^{2}-1}}^{2}\text{ \ \ \ }j=1,2,
\label{07}%
\end{equation}
implies that the norms $||r_{\rho_{d_{j}}}||_{\mathbb{R}^{d_{j}^{2}-1}},$
$j=1,2,$ of these Bloch vectors do not depend on what operator tuples
$\Upsilon_{d_{j}}$ are used in decomposition (\ref{05}). Moreover, from
relations (\ref{02}) and (\ref{07}) it follows that, for a pure bipartite
state $\rho_{d_{1}\times d_{2}},$ the norms of the Bloch vectors of the
reduced states under decompositions (\ref{05}) satisfy the relation
\begin{equation}
\frac{1}{d_{1}}+\frac{d_{1}-1}{d_{1}}||r_{\rho_{d_{1}}}||_{\mathbb{R}%
^{d_{1}^{2}-1}}^{2}=\frac{1}{d_{2}}+\frac{d_{2}-1}{d_{2}}||r_{\rho_{d_{2}}%
}||_{\mathbb{R}^{d_{2}^{2}-1}}^{2} \label{07_}%
\end{equation}
and
\begin{equation}
1-\mathrm{tr}(\rho_{d_{j}}^{2})=\frac{d_{j}-1}{d_{j}}\left(  1-||r_{\rho
_{d_{j}}}||_{\mathbb{R}^{d_{j}^{2}-1}}^{2}\right)  . \label{071}%
\end{equation}

Let $d_{k}=\min\{d_{1},d_{2}\}.$ Substituting relations (\ref{071}) into
formula (\ref{03}) specified for $j=k$, we have%
\begin{align}
\mathrm{C}_{\rho_{d_{1}\times d_{2}}}  &  =\sqrt{\alpha_{d_{1}\times d_{2}%
}\frac{d_{k}-1}{d_{k}}\left(  \text{ }1-\left\Vert r_{\rho_{d_{k}}}\right\Vert
_{\mathbb{R}^{d_{k}^{2}-1}}^{2}\right)  },\label{09}\\
d_{k}  &  =\min\{d_{1},d_{2}\}.\nonumber
\end{align}
Therefore, the normalization of the concurrence $\mathrm{C}_{\rho_{d_{1}\times
d_{2}}}$ for all $d_{1},d_{2}\geq2$ to the maximal value $1$, attained on a
maximally entangled state, i.e. $\left\Vert r_{\rho_{d_{k}}}\right\Vert
_{\mathbb{R}^{d_{k}^{2}-1}}=0,$ \ implies
\begin{equation}
\alpha_{d_{1}\times d_{2}}:=\frac{d_{k}}{d_{k}-1}\equiv\frac{\min\{d_{1}%
,d_{2}\}}{\min\{d_{1},d_{2}\}-1}. \label{010_}%
\end{equation}
Relations (\ref{01}) -- (\ref{010_}) prove the following general statement.

\begin{theorem}
Let $\rho_{d_{1}\times d_{2}},$ where $d_{1},d_{2}\geq2,$ be a pure bipartite
state on $\mathcal{H}_{d_{1}}\otimes\mathcal{H}_{d_{2}}$ and $\rho_{d_{j}},$
$j=1,2,$ be the states on $\mathcal{H}_{d_{j}},$ $j=1,2,$ reduced from
$\rho_{d_{1}\times d_{2}}$ and admitting representations (\ref{05}) with the
Bloch vectors
\begin{equation}
r_{\rho_{d_{j}}}=\sqrt{\frac{d_{j}}{2(d_{j}-1)}}\mathrm{tr}\left(  \rho
_{d_{j}}\Upsilon_{d_{j}}\right)  \in\mathbb{R}^{d_{j}^{2}-1},\text{
\ }\ j=1,2.\label{011}%
\end{equation}
The norms $\left\Vert r_{\rho_{dj}}\right\Vert _{\mathbb{R}^{d_{j}^{2}-1}}%
^{2}$,$j=1,2,$ of these Bloch vectors do not depend on a choice of operator
tuples $\Upsilon_{d_{j}}$ in representation (\ref{05}) and satisfy the
relation (\ref{07_}). The concurrence $\mathrm{C}_{\rho_{d_{1}\times d_{2}}}$
of a pure state $\rho_{d_{1}\times d_{2}}$ normalized to the maximal value
$1$, attained on a maximally entangled state, is given by
\begin{align}
\mathrm{C}_{\rho_{d_{1}\times d_{2}}} &  =\sqrt{1-\left\Vert r_{\rho_{d_{k}}%
}\right\Vert _{\mathbb{R}^{d_{k}^{2}-1}}^{2}},\label{013}\\
d_{k} &  :=\min\{d_{1},d_{2}\}.\nonumber
\end{align}

\end{theorem}

\section{Conclusions}

In the present article, we consistently develop the main issues of the Bloch
vectors formalism for a finite-dimensional quantum system. Within this
formalism, qudit states and their evolution in time, qudit observables and
their expectations, entanglement and nonlocality, etc. are expressed in terms
of vectors in the Euclidean space $\mathbb{R}^{d^{2}-1}$. Our developments
allow us:

\begin{itemize}
\item to formalize the main issues (Propositions 1--6, Corollary 1) of
decompositions of linear operators on a finite-dimensional complex Hilbert
space via different operator bases;

\item to find (Theorem 1) for all $d\geq2$ the new general expression for the
set of Bloch vectors of all traceless qudit observables and to describe the
properties of this set;

\item to find (Proposition 7, Corollary 2, Theorem 2) for the sets of Bloch
vectors of all qudit states, pure and mixed, the new compact general
expressions in terms of the operator norms, which have the unified form for
all $d\geq2$ and explicitly reveal the geometry properties of these sets. For
the sets of Bloch vectors under the generalized Gell-Mann representation these
properties cannot be analytically extracted from the known \cite{6,7}
equivalent specifications via systems (\ref{33}) of algebraic equations;

\item to derive (Theorems 3, 4) for all $d\geq2$ the new general equations for
the time evolution of the Bloch vector of a qudit state if a qudit system is
isolated and if it is open and to characterize (Propositions 8, 9) in both
cases the main properties of the Bloch vector evolution in time;

\item to express (Theorem 5) the concurrence of a pure bipartite state on
$\mathcal{H}_{d_{1}}\otimes\mathcal{H}_{d_{2}}$ of an arbitrary dimension
$d_{1}\times d_{2}$ via the norm of the Bloch vector of its reduced state
$\rho_{d_{k}}$ on the Hilbert space $\mathcal{H}_{d_{k}}$ of dimension
$d_{k}:=\min\{d_{1},d_{2}\}.$
\end{itemize}

The introduced general formalism is important both for the theoretical
analysis of quantum system properties and for quantum applications, in
particular, for optimal quantum control, since, for systems where states are
described by vectors in the Euclidean space, the methods of optimal control,
analytical and numerical, are well developed.

\section{Acknowledgments}

The study by E.R. Loubenets in Sections 2, 3, 4 of this work is supported by
the Russian Science Foundation under the Grant No 19-11-00086 and performed at
the Steklov Mathematical Institute of Russian Academy of Sciences. The study
by E.R. Loubenets and M. Kulakov in Sections 5, 6, 7 is performed at the
National Research University Higher School of Economics.

\end{document}